%% file: main.tex
\documentclass[journal]{IEEEtran}
\IEEEoverridecommandlockouts
\usepackage{cite}
\usepackage{amsmath,amssymb,amsfonts}
\usepackage{algorithmic}
\usepackage{graphicx}
\usepackage{textcomp}
\usepackage[table]{xcolor}
\usepackage{soul}
\usepackage{changepage}
\usepackage{multirow}
\usepackage[acronym]{glossaries}
\usepackage{circledsteps}
\usepackage{siunitx}
\usepackage[hidelinks]{hyperref}
\usepackage{booktabs}
\usepackage{cleveref}
\usepackage{enumitem}
\usepackage{listings}
\usepackage{subcaption}
\usepackage[normalem]{ulem}
\usepackage{orcidlink}
\def\BibTeX{{\rm B\kern-.05em{\sc i\kern-.025em b}\kern-.08em
T\kern-.1667em\lower.7ex\hbox{E}\kern-.125emX}}

\newcommand{\crossym}{\mbox{CrosSym}}
\newcommand{\hwklee}{\mbox{SEFOS}}
\newcommand{\systemc}{\mbox{SystemC}}
\newcommand{\riscv}{\mbox{RISC-V}}
\newcommand{\symsysc}{\mbox{SymSysC}}

\sethlcolor{cyan}

\begin{document}

    \newacronym{plic}{PLIC}{Platform Level Interrupt Controller}
    \newacronym{tlm}{TLM}{Transaction Level Modelling}
    \newacronym{rtl}{RTL}{Register Transfer Level}
    \newacronym{vp}{VP}{Virtual Prototype}
    \newacronym{smt}{SMT}{Satisfiability Modulo Theory}
    \newacronym{duv}{DUV}{Design under Verification}
    \newacronym{gcd}{GCD}{Greatest Common Divisor}
    \newacronym{bfs}{BFS}{Breadth First Search}
    \newacronym{mmio}{MMIO}{Memory Mapped Input/Output}

    \title{Comparing Methods for the Cross-Level Verification of SystemC Peripherals with Symbolic Execution
    \thanks{Received 03 July 2025; revised 03 November 2025; accepted 29 November 2025. This work was supported by the German Federal Ministry of Research, Technology, and Space under grant no.~16ME0127 and no.~16ME0135 (Scale4Edge), and grant no.~01IW24001 (EXCLplus) and no.~01IW25003 (ExaVerse). This article was recommended by Associate Editor W. Qian. \textit{(Corresponding author: Karl Aaron Rudkowski.)}}
    \thanks{Karl Aaron Rudkowski and Sallar Ahmadi-Pour are with the Institute of Computer Science, University of Bremen, 28359 Bremen, Germany (e-mail: karlaaron@uni-bremen.de; sallar@uni-bremen.de).}
    \thanks{Rolf Drechsler is with the Institute of Computer Science, University of Bremen, 28359 Bremen, Germany, and also with the Cyber-Physical Systems, DFKI GmbH, 28359 Bremen, Germany (e-mail: drechsler@uni-bremen.de).}
    \thanks{Digital Object Identifier 10.1109/TCAD.2025.3641038}
    }

    \author{Karl Aaron Rudkowski~\orcidlink{0009-0007-2035-7268}, Sallar Ahmadi-Pour~\orcidlink{0000-0003-4000-6207}, and Rolf Drechsler~\orcidlink{0000-0002-9872-1740}, \textit{Fellow, IEEE}}

\markboth{IEEE Transactions on Computer-Aided Design of Integrated Circuits and Systems}
{Rudkowski \MakeLowercase{\textit{et al.}}: Comparing Methods for the Cross-Level Verification of SystemC Peripherals with Symbolic Execution}

\IEEEpubid{\copyright 2025 The Authors. This work is licensed under a Creative Commons Attribution 4.0 License.}

    \maketitle

    \input{sections/00-abstract}


    \input{sections/01-introduction}
    \input{sections/03-preliminaries}
    \input{sections/02-related-work}
    \input{sections/04-main}
    \input{sections/05-eval}
    \input{sections/06-conclusion}

    \section*{Acknowledgment}
    The authors thank Marvin Bäcker for his support in preparing the evaluation.

    \bibliographystyle{IEEEtran}
    \scriptsize
    \bibliography{bib}

    \begin{IEEEbiography}[{\includegraphics[width=1in,height=1.25in,clip,keepaspectratio]{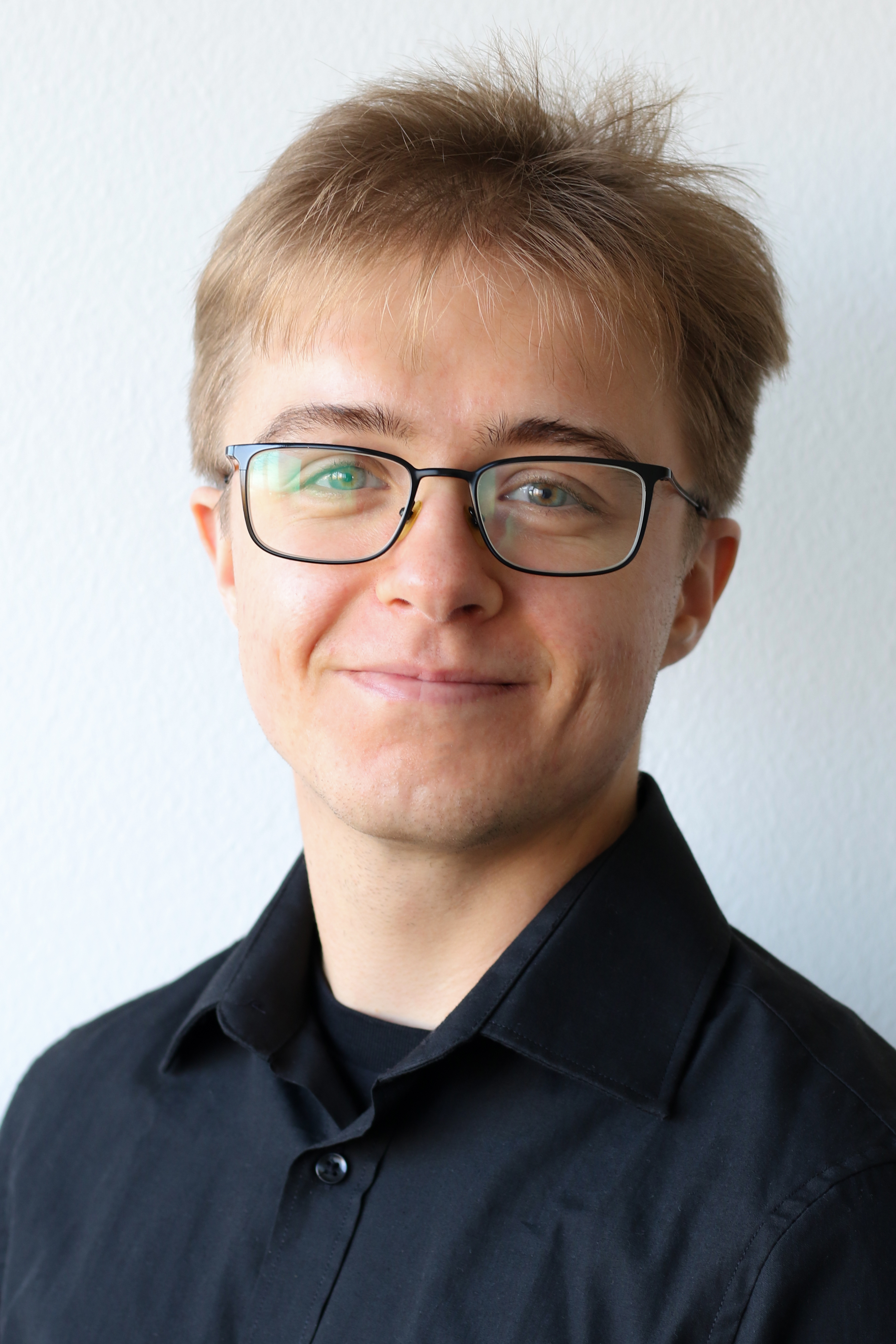}}]{Karl Aaron Rudkowski}
        received his B.Sc. and M.Sc. degree in Computer Science from the University of Bremen, Germany.
        Since 2024, he is pursuing the Ph.D. degree in the Group of Computer Architecture at the University of Bremen, Germany.
        His research interests include symbolic execution and its application to embedded systems.
    \end{IEEEbiography}

    \begin{IEEEbiography}[{\includegraphics[width=1in,height=1.25in,clip,keepaspectratio]{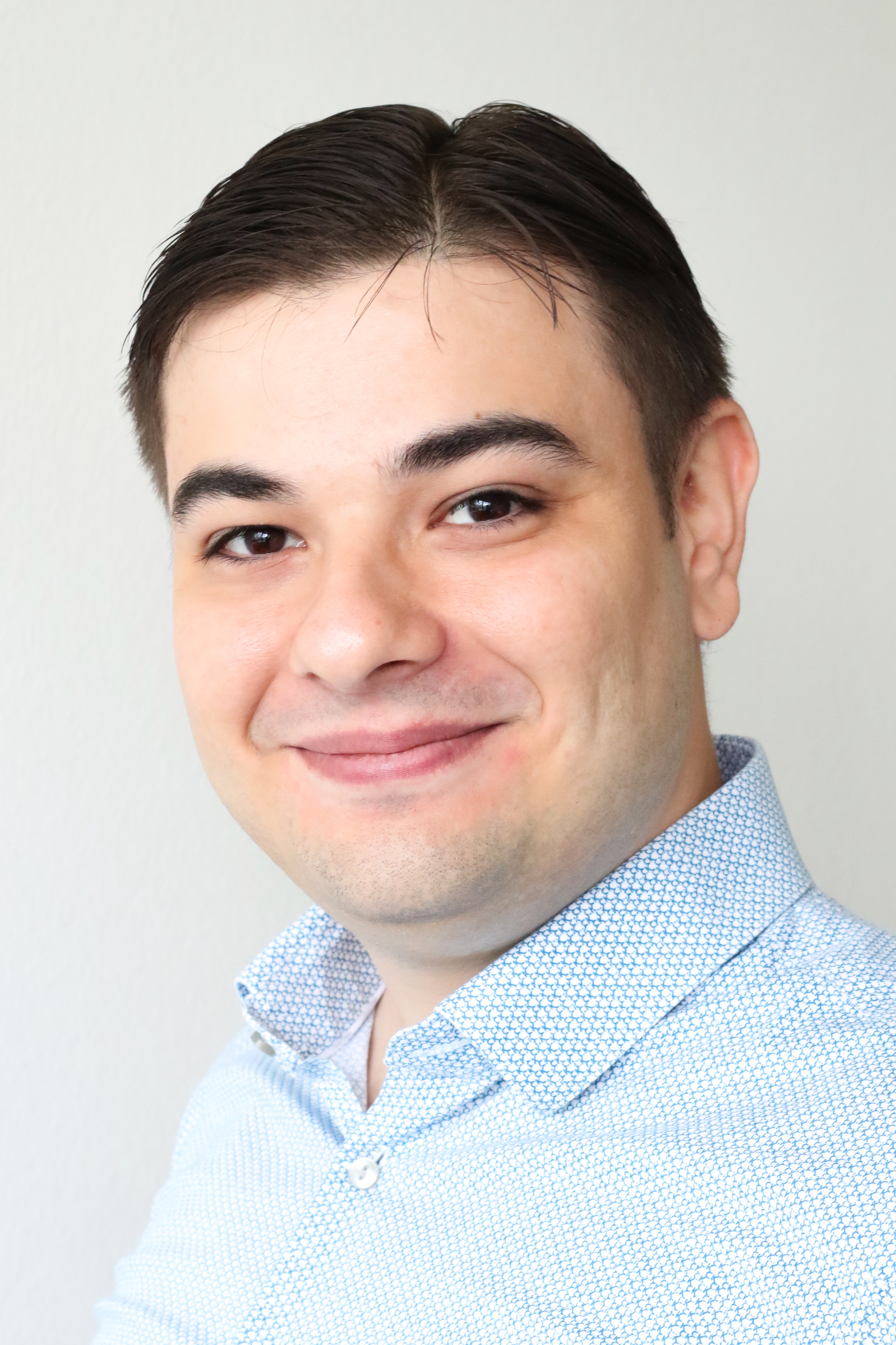}}]{Sallar Ahmadi-Pour}
        received his B.Sc. in computer engineering from the Hochschule Bremen University of Applied Sciences, Germany, in 2018, and his M.Sc. degree in computer science from the University of Bremen, Germany, in 2020.
        Since 2020 he has been pursuing his Ph.D. in the group of computer architecture at the University of Bremen, Germany.
        His research interests are the design and verification of computer architecture across different levels of system abstractions and the design of approximate circuits and systems.
    \end{IEEEbiography}

    \begin{IEEEbiography}[{\includegraphics[width=1in,height=1.25in,clip,keepaspectratio]{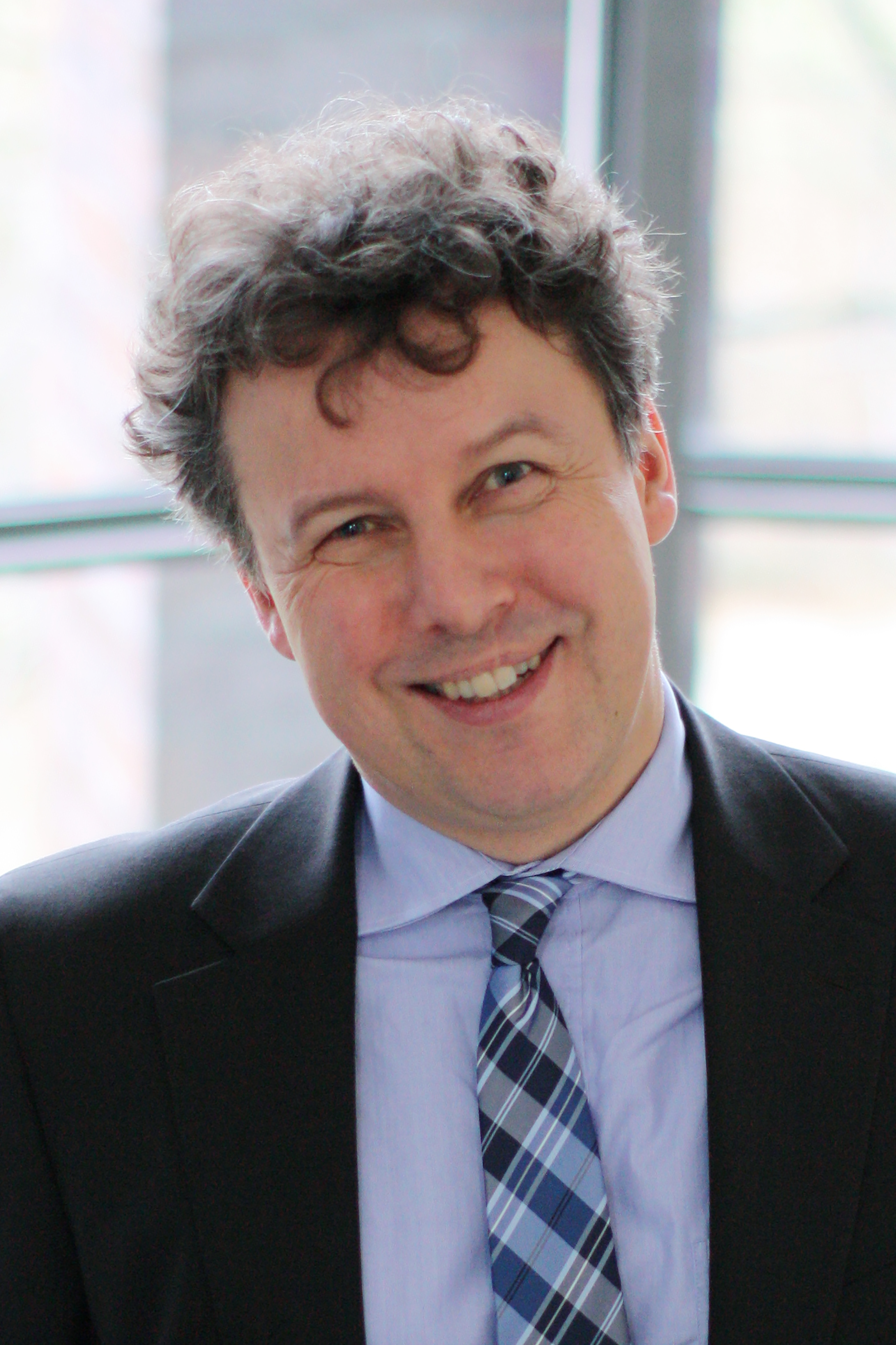}}]{Rolf Drechsler}
        received the Diploma and Dr.phil.nat. degrees in computer science from the Johann Wolfgang Goethe University in Frankfurt am Main, Germany, in 1992 and 1995, respectively.
        He worked at the Institute of Computer Science, Albert-Ludwigs University, Freiburg im Breisgau, Germany, from 1995 to 2000, and at the Corporate Technology Department, Siemens AG, Munich, Germany, from 2000 to 2001.
        Since October 2001, Rolf Drechsler is a Full Professor and Head of the Group of Computer Architecture, Institute of Computer Science, at the University of Bremen, Germany.
        In 2011, he additionally became the Director of the Cyber-Physical Systems Group at the German Research Center for Artificial Intelligence (DFKI) in Bremen.
        His current research interests include the development and design of data structures and algorithms with a focus on circuit and system design.
        He is an ACM Fellow and an IEEE Fellow.
        Prof. Drechsler was a member of Program Committees of numerous conferences including e.g., DAC, ICCAD, DATE, ASP-DAC, FDL, MEMOCODE, and FMCAD.
        He was Symposiums Chair at ISMVL 1999 and 2014, and the Topic Chair for ``Formal Verification'' at DATE 2004, DATE 2005, DAC 2010, and DAC 2011 and 2018.
        He was the General Chair of the IEEE European Test Symposium 2018 and the Program Chair of ICCAD 2020.
        He served as an Associate Editor of IEEE Transactions on Computer-Aided Design of Integrated Circuits and Systems, IEEE Transactions on Very Large Scale Integration Systems, Chip, IET Cyber-Physical Systems: Theory \& Applications, International Journal on Multiple-Valued Logic and Soft Computing, and ACM Journal on Emerging Technologies in Computing Systems.
    \end{IEEEbiography}

\end{document}

%% file: sections/00-abstract.tex
\begin{abstract}
\glspl{vp} are important tools in modern hardware development.
At high abstractions, they are often implemented in \systemc{} and offer early analysis of increasingly complex designs.
These complex designs often combine one or more processors, interconnects, and peripherals to perform tasks in hardware or interact with the environment.
Verifying these subsystems is a well-suited task for \glspl{vp}, as they allow reasoning across different abstraction levels.
While modern verification techniques like symbolic execution can be seamlessly integrated into \gls{vp}-based workflows, they require modifications in the \systemc{} kernel.
Hence, existing approaches modify and replace the \systemc{} kernel, or ignore the opportunity of cross-level scenarios completely, and would not allow focussing on special challenges of particular subsystems like peripherals.
We propose \crossym{} and \hwklee{}, two opposing approaches for a versatile symbolic execution of peripherals.
\crossym{} modifies the \systemc{} kernel, while \hwklee{} instead modifies a modern symbolic execution engine.
Our extensive evaluation applies our tools to various peripherals on different levels of abstractions.
Both tools' extensive sets of features are demonstrated for (1) different verification scenarios, and (2) identifying 300+ mutants.
In comparison with each other, \hwklee{} convinces with the unmodified \systemc{} kernel and peripheral, while \crossym{} offers slightly better runtime and memory usage.
In comparison to the state-of-the-art, that is limited to \gls{tlm}, our tools offered comparable runtime, while enabling cross-level verification with symbolic execution.
\end{abstract}

%% file: sections/01-introduction.tex
\section{Introduction}
In modern hardware design, \glspl{vp} are an important tool for dealing with the ever-increasing design complexity.
A \gls{vp} is an executable model of hardware which specifies the desired behaviour at a high level of abstraction.
It is commonly implemented in the C++-based \systemc{} library, using the \gls{tlm} extension~\cite{systemc}.
Such a hardware model consists of at least one processor, interconnects, and peripherals which provide an interface between the processor and the outside world.
With \glspl{vp}, verification and software design can start early on, long before any hardware exists~\cite{de2014better,herdt2021enhanced}.
Early and continuous verification is an important step, because errors found early are easier and cheaper to repair~\cite{de2014better}.
\IEEEpubidadjcol
To continuously confirm that refinement steps did not introduce any bugs, \textit{cross-level verification} can be applied.
This approach compares a model at a lower abstraction against a higher-level reference.
The verification method itself has to effectively search the state space, while seamlessly integrating into the development process.
One technique that offers both is \textit{Symbolic Execution}, which originates from the software domain.
It leverages symbolic variables, each representing a set of concrete values.
Thus, instead of a limited number of concrete test cases, it can offer a more complete reasoning about the design~\cite{symex-ueberblick}.
With regards to this verification, \systemc{} is interesting because it natively supports not just \gls{tlm}, but also models at lower abstraction levels such as the \gls{rtl}.
However, the original \systemc{} implementation poses challenges to modern symbolic execution engines, as discussed in \Cref{sec:crossym-journal:main:challenges}.
Due to this, current symbolic execution for hardware either (1) focuses exclusively on lower abstraction levels, avoiding \systemc{}, or (2) focuses on higher abstraction levels, but replaces \systemc{} with a constrained custom kernel.
With such a lack of versatility offered by the verification approaches, cross-level scenarios are not possible for any subsystems.
Additionally, the special challenges of peripherals are often not considered.
Unlike processors, peripherals implement a wide range of tasks.
Commonly used examples include communication protocols or sensors, while modern architectures have also brought forth more complex domain-specific applications~\cite{goldenage}.
Another concern is their communication interface, which allows them to interact with the system.
Both applications and interfaces mean that general or processor-specific verification approaches do not necessarily transfer to peripherals, and they therefore require separate attention.
To address this gap in cross-level symbolic execution of peripherals, we propose two tools, and accompanying considerations for their application.
In summary, our contributions are:
\begin{enumerate}
    \item Two opposing approaches to the challenges of symbolic \systemc{} execution, available open-source.
    \crossym~\footnote{\url{https://github.com/agra-uni-bremen/crossym}} modifies SystemC and peripheral to support the unmodified symbolic execution state-of-the-art.
    \hwklee~\footnote{\url{https://github.com/agra-uni-bremen/sefos}} (Symbolic Execution For Original SystemC) modifies the symbolic execution state-of-the-art to support the unmodified SystemC and peripheral.
    They are the first tools offering a comprehensive cross-level verification.
    \item Two propositions towards a more efficient verification.
    One considers how the peripheral communication interfaces influence the symbolic execution.
    The other targets the large object sizes in \systemc{}-based modelling.
    \item An examination of both our tools' applicability for a versatile verification as well as bug finding, by verifying four peripheral devices at \gls{rtl} and \gls{tlm}.
    The results additionally offer insight into how the fundamentally different approaches of \crossym{}  and \hwklee{} compare.
    \item An examination of both tools' performance in application to the S2C benchmark~\cite{s2c}, and in comparison to SymSysC~\cite{pascal-symsys} with regards to \gls{tlm} \systemc{}.
    SymSysC only supports this subset of features, and thus can only serve as a baseline, instead of a full comparison.
\end{enumerate}

\textbf{Paper Structure:}
This paper extends previous work of ours, namely \textit{\crossym{}: Cross-Level Verification of \systemc{} Peripherals Using Symbolic Execution}, which was published as a DDECS conference paper~\cite{crossym}.
In comparison with this preliminary version, our novel contributions are (1) \hwklee{} as a separate, alternative approach for cross-level symbolic execution, including an array minimization technique, (2) consideration of the peripheral communication interfaces, (3) extending the evaluation by additional peripherals, namely the Hash peripheral and the S2C benchmark, and (4) an extensive comparison between \crossym{} and \hwklee{} on the basis of our evaluation results.
The remainder of the paper follows with a discussion on relevant background information on \systemc{} and Symbolic Execution in \Cref{sec:crossym-journal:prelim}.
\Cref{sec:crossym-journal:rw} embeds our work into the context of related publications.
\Cref{sec:crossym-journal:main:challenges} presents the challenges that motivate our work, and gives a high-level overview of our contributions.
Following, \Cref{sec:crossym-journal:crossym} and \Cref{sec:crossym-journal:hwklee} introduce the two approaches \crossym{} and \hwklee{}, respectively.
\Cref{sec:crossym-journal:main:symsig} discusses how to handle the peripherals' bus interfaces in symbolic execution.
\Cref{sec:crossym-journal:eval} presents and discusses our evaluation, as well as the obtained results.
Finally, \Cref{sec:crossym-journal:ende} concludes our results and gives an overview over possible future research directions.

%% file: sections/03-preliminaries.tex
\section{Preliminaries}\label{sec:crossym-journal:prelim}
\textbf{\systemc:}
The \systemc~\cite{systemc} modelling language, based on C++, was initially developed for models at the \gls{rtl} abstraction level.
A model consists of modules encapsulating different components.
Their functionality is realised with processes, which can be either methods or threads, with the former being the go-to for \gls{rtl} models.
While methods are always executed in full and can be started an unlimited amount of times, threads are started only once and can be suspended before re-entering the previously saved context.
The simulation is event-driven, meaning the processes register themselves for execution on events such as clock edges, wait times, or incoming input.
After the scheduler is started, it continues to call processes and move forward the simulation time until there are no callable processes left, or the given time runs out.
At the \gls{rtl}, the modules communicate over signals and ports, and their operations are synchronised to clock edges.
These details come at the cost of a high simulation time.
Therefore the \gls{tlm} extension lowers the simulation time by abstracting central concepts.
The communication between modules uses simplified sockets to write and read.
Additionally, the simulation time is not advanced as strictly clock-cycle accurate as in \gls{rtl} modelling.
Processes can run (and advance their local simulation time) until pre-defined synchronization points are reached.
Here, threads are generally the processes of choice.

\textbf{Symbolic Execution:}
Symbolic Execution is a verification technique from the software domain.
There exist several state-of-the-art engines like KLEE~\cite{klee} and angr~\cite{angr}.
The idea is to represent sets of concrete values with symbolic variables~\cite{symex-ueberblick}.
If, during the execution, a symbolic variable is involved in an instruction, a \gls{smt} solver can be used to determine the instruction's result.
This solver can argue not only over boolean, but also over mathematical expressions consisting of multi-bit variables, arrays, bitvectors, and mathematical operators.
Branching instructions might fork the symbolic execution, and constrain the involved symbolic variables with new path conditions.
Others, such as memory accesses, division or assertions, can lead to errors.
For each possible error, the solver gives a concrete assignment to the symbolic variables.
Therefore, the program can be checked for generic failures, e.g. divide by zero, and violations of user-defined properties, expressed as assertions.
Modern engines follow a dynamic approach, where concrete and symbolic values are both possible.
If instructions only depend on concrete values, they are executed without the solver, and do not create new paths.
By defining which variables hold symbolic, and which concrete values, expert knowledge can guide the verification towards relevant paths.

%% file: sections/02-related-work.tex
\section{Related Work}\label{sec:crossym-journal:rw}
Prior work spans across a wide range of applied techniques.
Especially formal verification has been extensively studied.
This includes both model checking~\cite{herber-pockrandt-glesner,deng-wu-bian,bavonparadon-chongstitvatana}, and symbolic simulation~\cite{vladi-sym-tcad}.
However, these approaches require custom intermediate representations, which limits their applicability.
This inspired a transfer of techniques originating from the software domain, offering a more seamless integration into the workflow with \glspl{vp}.
One example is Cimatti et al.~\cite{sysc-sw}, who apply software model checking techniques to \systemc{} devices.
They tackle problems arising from the \systemc{} implementation, such as threading and inter-module communication.
Apart from formal methods, the software domain also offers simulation-based approaches that can offer a more efficient exploration.
One of these is coverage-guided fuzzing, which automatically generates inputs under the goal of increasing the coverage.
Hardware-oriented fuzzing applications include processors~\cite{niklas-fuzzing,instiller}, and peripherals in a cross-level scenario~\cite{sallar-crosslevel}.
Another method from the software domain is concolic testing, which combines fuzzing and symbolic execution.
Hardware-oriented concolic applications have been successful in testing a wide range of applications~\cite{ctsc,rtl-concolic,lyu-mishra,ahmed-farahmandi-mishra}.
However, both fuzzing and concolic are testing approaches, and cannot offer verification.
They are limited to the test cases they generate.
Among the simulation-based methods from the software domain, Symbolic Execution offers the possibility of a complete exploration.
Similar to symbolic simulation, concrete values are replaced with symbolic variables.
In contrast to symbolic simulation, which focuses on data flow, symbolic execution focuses on control flow~\cite{sylvia}.
Many hardware-oriented symbolic execution applications target \gls{rtl} Verilog models.
Sylvia~\cite{sylvia} was inspired by software symbolic execution engines, but is itself an independent engine.
While this allows for direct processing of Verilog code, such an engine also requires effort to reimplement fundamental techniques from the software domain, such as query caching~\cite{sylvia-query-caching}.
Therefore, most approaches translate Verilog models into C++ code, e.g. using Verilator~\cite{verilator}, and apply KLEE~\cite{klee}, the state-of-the-art symbolic execution engine for software.
SymbA~\cite{symba} and SE4RDV~\cite{zhang-feng-huang} apply this idea to general \gls{rtl} models.
Bruns et al.~\cite{niklas-processor} and Coppelia~\cite{coppelia} target \gls{rtl} processors.
While vanilla C++ is acceptable for \gls{rtl} models, it lacks the necessary constructs for \gls{tlm} models, which are usually implemented in \systemc{}.
Thus, these methods do not support a cross-level verification.
For \systemc{} implementations, one approach targets cross-level verification~\cite{crosslevel-symex}.
However, it compares \systemc{} implementations to UML specifications, and features a custom symbolic execution engine for \systemc{}.
To use the established state-of-the-art KLEE~\cite{klee}, SESC~\cite{lin-yang-cong-xie} and \symsysc~\cite{pascal-symsys} employ custom \systemc{} schedulers.
\mbox{SESC}~\cite{lin-yang-cong-xie} only supports the high-level synthesizable subset of \systemc{}.
Most importantly, events and \texttt{SC\_METHOD}s cannot appear in models.
\texttt{SC\_THREAD}s can only have static sensitivity to exactly one clock edge.
This heavily limits the models to which this approach can be applied.
\symsysc~\cite{pascal-symsys} is less limited, as \gls{tlm} features are generally supported, though \gls{rtl} features are not.
Additionally, \symsysc{} explicitly verifies peripherals.
In conclusion, symbolic execution offers the possibility for verification while avoiding custom intermediate representations.
However, most symbolic execution based approaches are restricted in the devices and abstraction levels to which they are applicable.
The specific challenges of peripherals are usually not considered.
While \symsysc{} is closest to our goal of verifying \gls{rtl} and \gls{tlm} peripherals, it does not support \gls{rtl} peripherals.
However, to the best of our knowledge, there are no other approaches which achieve our goal.

%% file: sections/04-main.tex
\section{Symbolically Executing \systemc{}:\\ Challenges and Approaches}\label{sec:crossym-journal:main:challenges}
When verifying \systemc{} peripherals, the implementation of the original \systemc{} kernel causes three main obstacles.
\begin{enumerate}
    \item The \texttt{SC\_THREAD}s, commonly used in \gls{tlm} models, are realised using system threads.
    While there are approaches for the symbolic execution of multi-threaded software~\cite{cloud9}, this is currently not supported by state-of-the-art tools due to the associated large overhead.
    \item The signal-port mechanism, commonly used for inter-module communication in \gls{rtl} models, uses \texttt{mutexes}, which are likewise not supported.
    \item The long list of supported features naturally leads to an implementation with many branch instructions and large class object sizes.
    Both put a strain on the \gls{smt} solver, and thus slow down the verification, possibly to the point of rendering it ineffective.
\end{enumerate}
Therefore, the type of verifiable peripherals is heavily constrained.
To combat these issues, two approaches for the cross-level symbolic execution are conceivable:
\begin{table}[h!]
    \centering
    \resizebox{\linewidth}{!}{
        \begin{tabular}{ccc}
            \toprule
            &\textbf{\systemc{} Implementation} & \textbf{Symbolic Execution Engine} \\
            \textbf{I)}&\cellcolor{orange!50}Modified & \cellcolor{cyan!50}Original \\
            \textbf{II)}&\cellcolor{cyan!50}Original & \cellcolor{orange!50}Modified \\
            \bottomrule
        \end{tabular}
    }
\end{table}

Either the established symbolic execution tool is employed as-is, and the \systemc{} kernel as well as \gls{duv} are modified (I).
While this is the state-of-the-art for \systemc{} peripherals~\cite{pascal-symsys}, the supported features are inherently limited.
As such, a cross-level verification is currently not possible.
In addition, the equivalence to the original implementations is not proven, and the idea often cannot be integrated into industry workflows.
An alternative consists of supporting the \textit{unmodified} \systemc{} kernel and \gls{duv} (II).
Instead, small modifications in the established symbolic execution engine specifically target \systemc{}.
This constrained domain allows optimizations which would not be possible for the verification of general software.
Consequently, one benefits from the symbolic execution state-of-the-art and the unmodified \systemc{} and \gls{duv} implementations.
These two approaches result in two opposing tools, which will be presented in the following.
\crossym{} implements (I), revolving around a replacement for the \systemc{} kernel.
\hwklee{} (Symbolic Execution For Original SystemC) implements (II), revolving around small modifications to the state-of-the-art symbolic execution engine.

\subsection{Verification Workflow}
\begin{figure}[h]
    \centering\includegraphics[width=0.80\linewidth]{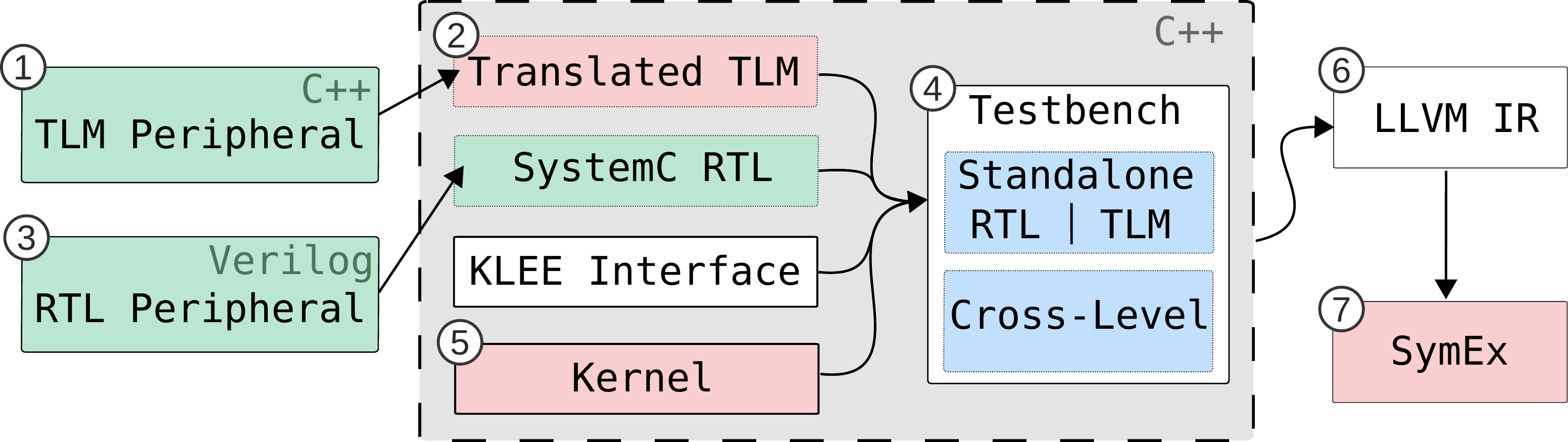}
    \caption{Verification Workflow\label{fig:crossym-journal:workflow}}
\end{figure}
In \cref{fig:crossym-journal:workflow}, we give an overview of the proposed verification workflow.
Most of it is identical for both approaches.
The desired \gls{duv} can be provided at both the \gls{rtl} and \gls{tlm} level.
The \gls{tlm} description~\Circled{1} is usually readily available in \systemc{}.
Similarly, the \gls{rtl} description~\Circled{3} of the \gls{duv} has to be available in \systemc{}, which would most commonly be generated from Verilog using Verilator~\cite{verilator}.
Such accessible open source tools, as well as \systemc{}'s native support of both abstraction levels, make this extra step worth considering.
A testbench~\Circled{4} defines the desired behaviour of the \gls{duv} using assertion statements, and declares symbolic variables as well as assumptions with interface methods of the symbolic execution engine.
Using concrete values for all non-symbolic variables means that a test case only has to explore the corresponding relevant parts of the \systemc{} kernel.
By defining the desired interactions and simulation times in a test case, the symbolic execution can be guided towards, and limited to, relevant paths.
Thus, unbounded loops inside the \gls{duv}, such as those waiting for signal input or status changes, or processing input, can be bound to user-defined limits.
The tests can realise three possible verification scenarios: (a) Standalone verification of a \gls{tlm} peripheral, (b) Standalone verification of an \gls{rtl} peripheral, and (c)  Cross-level verification \gls{rtl}$\Leftrightarrow$\gls{tlm}.
If both \gls{rtl} and \gls{tlm} implementations of a peripheral are available, the testbench might of course cover all three scenarios in separate tests.
The cross-level scenario benefits from a simplified test development at the \gls{rtl}.
Given that the desired behaviour is already specified, most of a test case boils down to a direct comparison, instead of rewriting assertions.
This process is less time-consuming and error-prone.
The testbench is compiled into the LLVM \textit{Intermediate Representation}~\Circled{6} alongside the \systemc{} peripheral(s) and kernel~\Circled{5}.
Finally, the peripheral can be verified using the relevant symbolic execution engine.
In \cref{fig:crossym-journal:workflow}, the three red boxes mark details that change depending on the approach.
The threads in \gls{tlm} peripherals need to be translated~\Circled{2} for \crossym{} as described in~\cite{pascal-symsys}.
This step is not necessary for \hwklee{}, as it supports the unmodified \gls{duv} and \systemc{}.
The kernel~\Circled{5} is either a replacement implementation (for \crossym{}) or the original (for \hwklee{}).
The symbolic execution engine~\Circled{7} is either KLEE (for \crossym{}) or modified KLEE (for \hwklee{}).

This workflow can be considered in parallel to alternative verification and testing approaches.
The necessary annotations in the testbench are minimal, as only input needs to be declared symbolic.
For \crossym{}, one limiting factor are the thread transformations, which constitute changes in the \gls{duv}.

\subsection{Application-Specific Optimization}\label{sec:crossym-journal:main:app-op}
The original \systemc{} kernel supports many concepts for a wide range of applications.
However, this complexity can negatively impact the verification time, which is already a concern in comparison to testing approaches such as fuzzing.
An example of such a feature is asynchronous waiting, which introduces \texttt{mutexes} into the signal-port mechanism.
This enables the communication between different \systemc{} simulators, or between a simulator and the operating system.
However, for our goal of verifying peripherals in isolation, the set of necessary features is smaller.
Asynchronous waiting, for example, is not necessary, since we do not aim to verify entire systems or interactions between multiple peripherals.
Including it, and similar unused features, would disadvantage the symbolic execution, possibly to the point of rendering it ineffective.
Excluding it is simple, even when integrating the original \systemc{} kernel, with the help of a build option.
In the replacement kernel, we have even greater freedom in choosing a subset of features.
As a consequence, we carefully chose the supported features to cover our use case.

\subsection{Feature Comparison}
As discussed above, replacements of the \systemc{} kernel implement a limited subset of features.
To illustrate the different scopes of both our tools and the state-of-the-art~\cite{pascal-symsys}, \Cref{tab:crossym-journal:comp} shows a comparison between their supported features.
\begin{table}
    \centering
    \caption{\systemc{} Feature Comparison Between \symsysc{} and our Tools (Orange Checkmark = Limited Support)}.\label{tab:crossym-journal:comp}
    \resizebox{\linewidth}{!}{
    \begin{tabular}{ccccc}
        \toprule
        \multicolumn{2}{c}{\multirow{2}{*}{\textbf{Feature}}} & \multirow{2}{*}{\textbf{\symsysc~\cite{pascal-symsys}}} & \multicolumn{2}{c}{\textbf{Our Work}} \\
        & & & \textbf{\crossym{}} & \textbf{\hwklee{}} \\
        \hline
        \parbox[t]{1mm}{\multirow{5}{*}{\rotatebox[origin=c]{90}{\textbf{TLM}}}}
        & Threads & \color{orange} Modified \gls{duv} & \color{orange} Modified \gls{duv} &\color{blue!20!black!30!green} \checkmark \\
        & Events & \color{orange}(\checkmark) & \color{blue!20!black!30!green} \checkmark &\color{blue!20!black!30!green} \checkmark \\
        & \texttt{wait(event)} & \color{orange} (\checkmark) & \color{blue!20!black!30!green} \checkmark &\color{blue!20!black!30!green} \checkmark \\
        & Time & \color{blue!20!black!30!green} \checkmark & \color{blue!20!black!30!green} \checkmark &\color{blue!20!black!30!green} \checkmark \\
        & \texttt{wait(time)} & \color{blue!20!black!30!green} \checkmark & \color{blue!20!black!30!green} \checkmark &\color{blue!20!black!30!green} \checkmark \\
        \hline
        \parbox[t]{1mm}{\multirow{5}{*}{\rotatebox[origin=c]{90}{\textbf{RTL}}}}
        & Methods & \color{blue!20!black!30!green} \checkmark & \color{blue!20!black!30!green} \checkmark & \color{blue!20!black!30!green} \checkmark \\
        & Static sensitivity &\color{red} $\times$ & \color{blue!20!black!30!green} \checkmark &\color{blue!20!black!30!green} \checkmark \\
        & Signals \& Ports &\color{red} $\times$ & \color{blue!20!black!30!green} \checkmark &\color{blue!20!black!30!green} \checkmark \\
        & Delta cycles &\color{red} $\times$ & \color{blue!20!black!30!green} \checkmark &\color{blue!20!black!30!green} \checkmark \\
        & Clock & \color{red} $\times$ & \color{blue!20!black!30!green} \checkmark &\color{blue!20!black!30!green} \checkmark \\
        \hline
        \multicolumn{2}{c}{Other \systemc{} features} & \color{red} $\times$ & \color{red} $\times$ & \color{blue!20!black!30!green}\checkmark \\
        \bottomrule
    \end{tabular}
    }
\end{table}
The first column lists the most important \gls{rtl} and \gls{tlm} features.
While the features can be used at both abstraction levels, we listed them with the most relevant one.
The last three columns show the supported features in the state-of-the-art \symsysc~\cite{pascal-symsys}, and our tools \crossym{} and \hwklee{}.
An orange checkmark signifies basic support with limitations.
\symsysc{} and \crossym{} are based on the same idea of replacing the \systemc{} kernel.
Therefore, as denoted in the last line, they are limited to the features explicitly listed in the table, while \hwklee{} in theory supports most of the original \systemc{} features.
Nonetheless, in the context of this paper, we focus on those that demonstrably work, according to our evaluation.
In addition to this limitation, \symsysc{} and \crossym{} are also nearly identical for \gls{tlm}.
For \texttt{SC\_THREAD}s, they require modifications of the \gls{duv}, which are not necessary for \hwklee{}.
However, implementation details of the replacement kernels differ.
\symsysc{}'s event support is limited, as only one thread at a time can wait for an event.
Additionally, threads that call \texttt{wait(event)} after an \texttt{event.notify(time)} call, but before the actual notification time, are not notified, which violates the \systemc{} standard~\cite{systemc}. % 5.10.6: notify and cancel
Both of these issues are not present in \crossym{}.
Further, \crossym{} supports the \gls{rtl} features.
In the following, we present the necessary considerations for a cross-level symbolic execution of \systemc{} peripherals.
This mainly consists of our two opposing tools, \crossym{} (\Cref{sec:crossym-journal:crossym}) and \hwklee{} (\Cref{sec:crossym-journal:hwklee}), each implementing one of the discussed ideas.
Additionally, application-specific optimizations ensure reasonable runtimes specifically for peripherals (\Cref{sec:crossym-journal:main:symsig}).

\section{\crossym{}: Cross-level Verification Using a Replacement Kernel}\label{sec:crossym-journal:crossym}
The original \systemc{} kernel poses multiple challenges to modern symbolic execution tools.
These make an alternative kernel implementation, written specifically with symbolic execution in mind, a solution worth considering.
Such a replacement kernel was proposed for example by Pieper et al~\cite{pascal-symsys}, who focused on \gls{tlm} peripherals.
Noteworthy is the thread translation, which is part of addressing the problems caused by system threads.
The peripherals' \texttt{SC\_THREAD}-function code is transformed, so that the expected control flow is respected without the need for thread-based context switches.
However, the kernel was limited to \gls{tlm} features, and misses all \gls{rtl} features.
The authors do not address cross-level verification.

\subsection{Replacement Kernel}
Our kernel supports the \gls{tlm} concepts of \symsysc{} with improvements in the event notification system.
However, its main benefit are the added \gls{rtl} concepts of static sensitivity, the clock, and signals and ports as well as the delta cycles to mimic their value updates.
These make the verification of \gls{rtl} peripherals possible.
Processes can be made reactive to event notifications or signal/port value changes during module instantiation.
Each event/signal/port entity collects its reactive processes locally.
If an entity has a update during the simulation, the update is executed in the corresponding time step.
For this, the scheduler retrieves the entity's waiting processes, and executes them.
Signals and ports transmit values in and out of modules, and are therefore heavily intertwined.
The clock is a signal which continuously writes its own value according to the duty cycle specified by the user.
To ensure that processes always read the correct signal update, each time step simulation is split into the \textit{evaluation-update-delta} cycles known from the original \systemc~\cite{systemc}.
The above-mentioned improvement to \symsysc's~\cite{pascal-symsys} event notification system stems from (1) the events collecting all waiting processes (instead of only one), and (2) the scheduler's retrieval of waiting processes from the event at time of update, instead of at time of notification.

\section{\hwklee{}: Cross-level Verification Using the Original \systemc{} Kernel}\label{sec:crossym-journal:hwklee}
While the replacement kernel offers an opportunity for optimization, it constitutes an interference in the peripheral.
This is even more obvious for the necessary \texttt{SC\_THREAD} transformation, which directly modifies the code and requires user effort for every \gls{duv}.
An alternative verification approach is a symbolic execution engine that accommodates the problems mentioned in \Cref{sec:crossym-journal:main:challenges}.
It thereby supports unmodified peripherals, and the original \systemc{} implementation.
Such an engine is faced with a different problem, i.e. the overhead associated with verifying multi-threaded software, such as checks for data races.
However, we focus specifically on hardware models implemented in \systemc{}, which offers some assumptions to rely upon.
The \systemc{} standard~\cite{systemc} specifies \textit{cooperative multitasking}.
That means that all processes, including \texttt{SC\_THREAD}s, are executed without interruption, and sequentially.
The developer defines synchronization points, e.g. with \texttt{wait()} statements.
System threads enable the context switches between \texttt{SC\_THREAD}s, but they do not execute in parallel.
Therefore, many threading issues are not a topic for the symbolic execution of \systemc{} peripherals.
Instead, the clearly defined process-switch protocol can be embedded directly into the engine.
Thus, an exploration of the peripherals becomes possible.

\subsection{Threading in \systemc{}}
\begin{figure}
    \centering\includegraphics[width=0.8\linewidth]{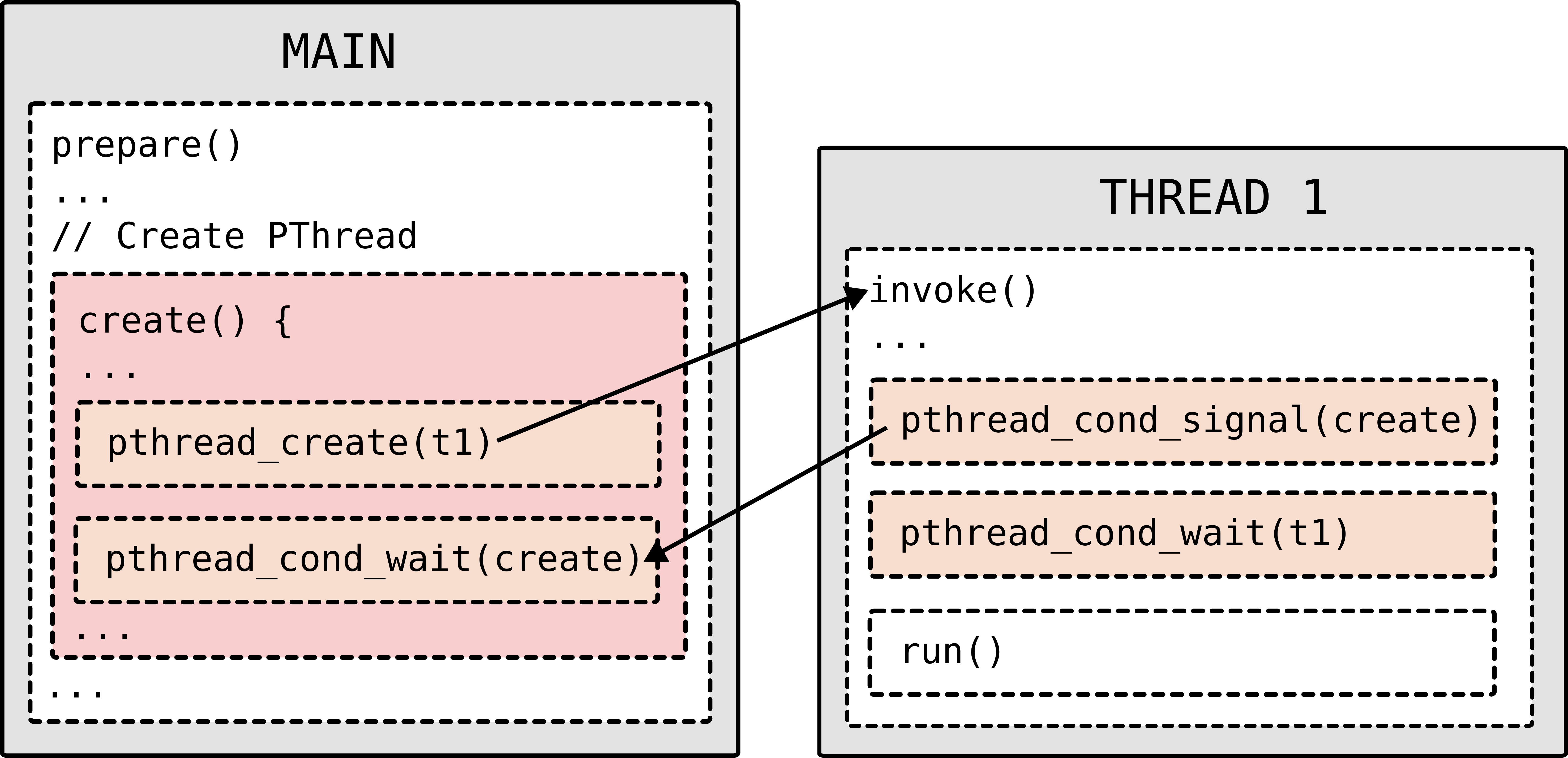}
    \caption{\systemc{} Thread Creation (PThread module)\label{fig:crossym-journal:symex:create}}
\end{figure}

\systemc{} offers multiple implementations of the threading behaviour defined in the standard.
These differ in the threading library they are based on, but the general attributes remain the same.
We focused on the PThread-based version.
In it, each \texttt{SC\_THREAD}, as well as the scheduler, is its own PThread, associated with a unique \texttt{pthread\_condition} (in the following \texttt{pt\_cond}).
All threads are created before the simulation start, as illustrated in \cref{fig:crossym-journal:symex:create}.
A special \texttt{pt\_cond create} ensures that the scheduler initialises the PThreads one after the other.
During the setup, the new thread does not execute the \texttt{SC\_THREAD}-declared function (here: \texttt{run}).
Instead, it starts waiting on its newly created \texttt{pt\_cond}.
\begin{figure}
    \centering\includegraphics[width=\linewidth]{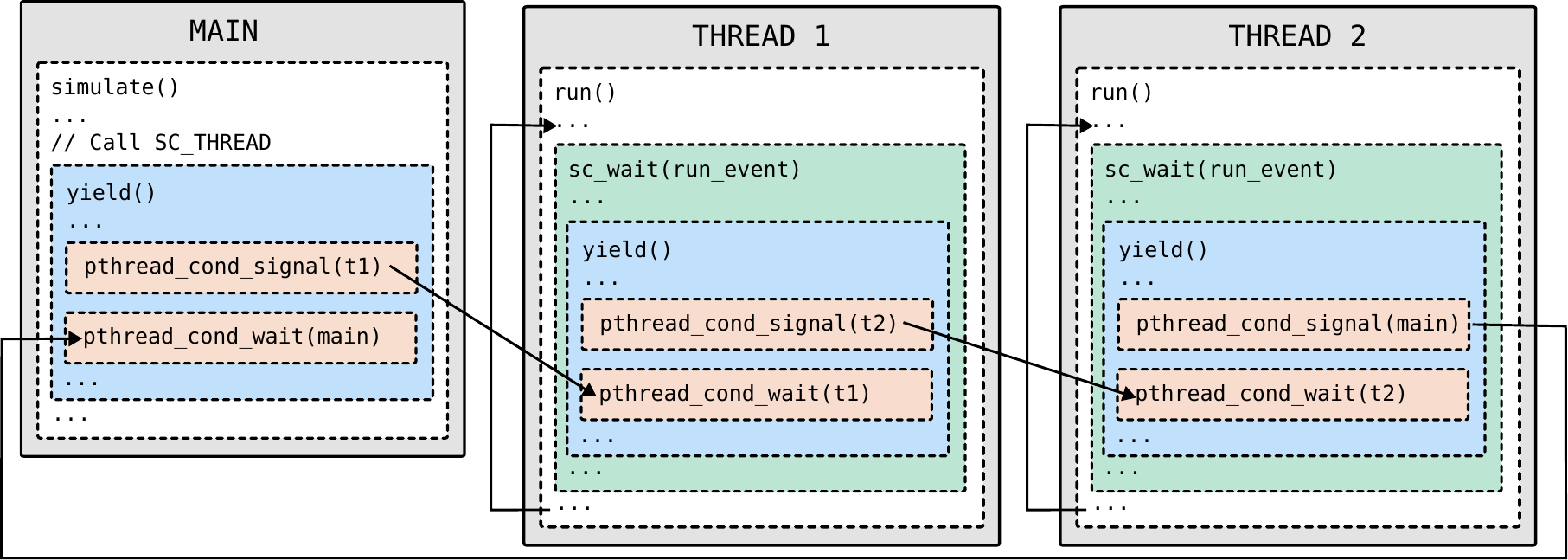}
    \caption{\systemc{} Threading (PThread module)\label{fig:crossym-journal:symex:threading}}
\end{figure}
During the simulation, the thread execution is managed over a \texttt{pt\_signal}/\texttt{\_wait} handshake, as illustrated in \cref{fig:crossym-journal:symex:threading}.
To switch from one thread to another, the callee thread first signals the called thread's \texttt{pt\_cond}.
Then, it starts waiting on its own \texttt{pt\_cond}, and only continues execution after it is signalled again.
Therefore, at any time, only one thread is actually executing.

\subsection{Threading in Symbolic Execution of \systemc{}}
\begin{figure}
    \centering\includegraphics[width=\linewidth]{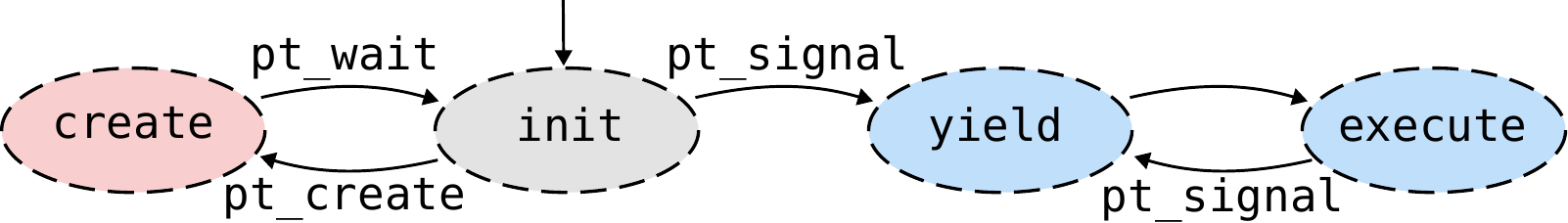}
    \caption{Symbolic Execution Engine States\label{fig:crossym-journal:states}}
\end{figure}
With the assumptions defined in the \systemc{} standard~\cite{systemc} and illustrated in the previous section, the necessary threading support of the symbolic execution engine can be simplified.
Our engine is based on KLEE~\cite{klee} version 3.1.
In it, each thread needs its own context, represented by stack and program counter.
These save the relevant information to recover the thread's context after a switch.
The stack is initially empty, and contains only variables allocated while the respective thread is active.
During the symbolic execution, the \texttt{CallInst}s targeting \texttt{pt\_*} functions are intercepted.
Their original implementation is ignored, which avoids the creation of real system threads.
Instead, they mark an internal state change, as illustrated in \cref{fig:crossym-journal:states}.
In \textit{init}, the main thread's context is active.
In \textit{create}, \texttt{sysc::invoke} is executed in a new context, as seen in \cref{fig:crossym-journal:symex:create}.
After \texttt{pt\_wait}, this context is associated with the new \texttt{pt\_cond}.
Back in \textit{init}, the main context continues with the instruction after \texttt{pt\_create}.
The first \texttt{pt\_signal} triggers \textit{yield}, as shown in \cref{fig:crossym-journal:symex:threading}.
During \textit{yield}, the current context is saved, and the desired one is loaded.
Afterwards, regular \textit{execution} continues.
If the execution has to be forked, both resulting states inherit the context information for all active threads.
The states then continue thread execution independent of each other.
Outside of the three \texttt{pt\_*} \texttt{CallInst}, all LLVM instructions are handled as specified.
Thus, the original implementations of \systemc{} and \gls{duv} are considered.
This stands in contrast with \crossym{} and the current state-of-the-art~\cite{pascal-symsys}, where both \systemc{} and \gls{duv} are modified.
While this approach relies on the correctness of the \systemc{} kernel's threading implementation, it is less vulnerable than a completely new kernel implementation.
Additionally, all other parts of the kernel which are encountered during the execution are still checked.
This offers more safety than concretising all input handed to the kernel, as offered e.g. by S2E~\cite{s2e}.

\subsection{Array Minimization}\label{sec:crossym-journal:main:array-min}
The previous sections discussed the effort required to start the symbolic execution of unmodified \systemc{} peripherals.
However, during the runtime, another issue arises, namely the predominance and size of class objects.
Each \gls{duv} and testbench is implemented as a C++ class, and they integrate a number of \systemc{} features that are themselves implemented as C++ classes.
Whenever the \gls{duv} is relevant in a query, the symbolic execution engine embeds the object state into the query.
It is represented by an array, and a list of value updates.
However, the larger an object is, and therefore the corresponding array, the longer a query takes.
Consequently, a considerable amount of the symbolic execution time is spent on resolving accesses into objects.
Only specific cases can avoid these expensive solver queries.
Examples include cases where the object is accessed with a concrete index, and has only ever been written to using concrete indices.
KLEE~\cite{klee} already catches these base cases.
In any cases where multiple values are possible, the entire object is embedded in the array.
Examples for this include cases where a range of object indices is possible, or multiple updates with symbolic indices exist.
This problem can be addressed in different ways.
First, the C++ object size itself can be minimised as much as possible.
Such an approach is possible for \crossym{}, where only the necessary subset of features is implemented.
However, for \hwklee{}, where the original \systemc{} implementation is integrated, this is not possible to any practical degree.
While a few features can be deactivated using build options, the overwhelming majority cannot.
A second approach that already exists in the literature is a fundamental rework of how allocated objects are internally managed~\cite{memobj-alternative}.
In this scenario, an object can be split up into $x$ smaller objects.
For a symbolic access, the execution has to fork, creating up to $x-1$ new states.
However, both the fundamental changes and the forking make alternative approaches worth considering.
Our goal is to make symbolic execution of \systemc{} feasible with minimal changes to the symbolic execution engine.
Additional, the state-space explosion, i.e. creating an infeasible number of states during the execution, is already a common problem of symbolic execution.
We prefer to avoid any further forking whenever possible.
Finally, a third approach minimises not the C++ objects, but the arrays representing these objects.
This is only possible for \hwklee{}, because modifications to the symbolic execution engine are necessary.
\crossym{}, however, is based around the idea of applying unmodified symbolic execution.
For most queries, only parts of an object are actually relevant.
Consider an object $O$ which is accessed using address $\alpha$.
On a specific path, $\alpha$ might refer to any of $O$'s input registers, but never to an output register.
In this case, the output registers are not relevant, and do not have to be represented in any arrays sent to the solver.
Consequently, for the result $O[\alpha]$ that will be used in future queries, the relevant object parts should be identified.
``Relevant'' here refers to an approximation of $O$, that contains at minimum all possible $O[\alpha]$.
The more impossible $O[\alpha]$ are filtered out, the better.
However, this filtering requires solver queries, therefore consuming resources.
Even though one numerical query such as $\alpha==1$ is generally faster than one array query, a high amount of numerical queries could still result in an overall worse runtime.
Therefore, we avoid an exact resolution of $O$ that contains \textit{only} possible $O[\alpha]$.
The actual filtering targets the two components $O$ is represented by, the state array $s$ and the update list $u$.
Both $s$ and $u$ can contain concrete and symbolic values.
The difference is that all $s_y \in s$ were written with concrete indices, while all $u_x \in u$ were written with symbolic indices.
The access $O[\alpha]$ results in the approximation $O[\alpha]=s_{min},u_{min}$.
To identify $s_{min}$, the solver has to be queried for $\alpha_{min}$ and $\alpha_{max}$, the extreme concrete assignments of $\alpha$.
The final, shortened array $s_{min}$ consists of all $s[i]$ where
\begin{align*}
    & i \geq \alpha_{min} \land i \leq \alpha_{max} \\
    \land & i \geq s.address \land i \leq (s.address+s.size)
\end{align*}
This corresponds at minimum to all valid $s$ entries addressable by $\alpha$, because $\alpha$ cannot take on values smaller than $\alpha_{min}$ or greater than $\alpha_{max}$.
Therefore, our most important requirement is met.
No possible $s[\alpha]$ values are eliminated by our preprocessing.
However, it is possible that some $i \in s_{min}$ are not valid assignments for $\alpha$, given current path constraints.
For example, consider $\alpha_{min}=3, \alpha_{max}=6,\alpha\neq5$.
$s_{min}$ contains $s[5]$, even though $\alpha=5$ is not possible.
This is also in line with our previously mentioned requirements, because detecting $s[5]$ would require unwanted additional solver time.
To identify the relevant updates $u_{min}$, the solver has to solve $\alpha == u_i$ for each symbolic indices.
Here, we approximate by querying if $\alpha == u_i$ might be true, not if it has to be.
If an update $u_i$ might be relevant, it is included in $u_{min}$.
Similarly to $s_{min}$, $u_{min}$ will thereby contain at minimum all possible $u[\alpha]$, and maybe some impossible ones.

For the algorithmic complexity of the array minimization, both the \gls{smt} solver queries and the value processing are relevant.
The processing takes $O(n*m)$ for reading $n$ bytes from an object with $m$ updates.
For each byte, each update requires an \gls{smt} solver query.
To find $\alpha_{min/max}$ for the 32 bit wide $\alpha$, a constant amount of solver queries is necessary.
In both cases, these queries can take exponential time because \gls{smt} solving is generally a NP-hard problem.
However, both KLEE~\cite{klee} and modern \gls{smt} solvers preprocess the queries with a wide range of techniques, such as expression rewriting or caching, which can improve the performance drastically~\cite{klee}.

\section{Communication Interfaces}\label{sec:crossym-journal:main:symsig}
Regardless of the tool and the integrated \systemc{} kernel, peripherals pose one additional challenge.
They are connected to the system with communication interfaces.
This interaction introduces an additional layer of complexity to the verification.
\input{sections/05-eval-peripheral}
Symbolic values first have to pass through the interface logic before reaching any functionality.
However, there is generally little correlation between the general, standardised communication, and the peripheral-specific functionality code.
Therefore, verifying these two subtasks in a single test case can negatively impact the results for both.
\begin{figure}
    \centering
    \begin{subfigure}[t]{0.48\linewidth}
        \centering
        \subcaption{Write Test Input}\label{fig:crossym-journal:symbolicsignals:test}
        \begin{lstlisting}[breaklines=true]
a = klee_int("a");
in.write(a);
b = klee_int("b");
in.write(b);
        \end{lstlisting}
    \end{subfigure}
    \begin{subfigure}[t]{0.48\linewidth}
        \centering
        \subcaption{\systemc{} Implementation}\label{fig:crossym-journal:symbolicsignals:sysc}
        \begin{lstlisting}[breaklines=true]
write(new):
    if(val!=new)
        val=new;
    else return;
        \end{lstlisting}
    \end{subfigure}
    \caption{Interaction of Symbolic Inputs and \systemc{} Signals}\label{fig:crossym-journal:symbolicsignals}
\end{figure}
This effect can be observed most clearly for \gls{rtl} peripherals, which communicate over signals and ports as illustrated in \cref{fig:crossym-journal:symbolicsignals}
Fig.~\ref{fig:crossym-journal:symbolicsignals:test} shows an excerpt from a testbench, where inputs \texttt{a} and \texttt{b} are written over the same bus, and therefore the same signal \texttt{in}.
However, \systemc{} specifies that a signal is only updated if the new and old value differ~\cite{systemc}.
Therefore, the new and old have to be compared, as shown in \cref{fig:crossym-journal:symbolicsignals:sysc}.
Writing \texttt{a} triggers a comparison to the concrete initial value \texttt{c}, and thus paths \texttt{a=c} and \texttt{a!=c}.
Writing \texttt{b} creates the paths \texttt{a=b} and \texttt{a!=b} for both existing paths, resulting in $4$ paths overall.
Generally, this number depends on the number of signals, ports, and symbolic inputs.
For each of these paths, the entire functionality has to be explored.
This can lead to a drastically higher runtime, in extreme cases going from a few minutes to over \qty{24}{\hour}.
As a consequence, we split the verification of communication and functionality into separate test cases.
Special constraints on inputs or outputs can be specified in the test setup itself.
For the communication, only the post-transaction register state has to be checked.
For the functionality, symbolic values can be written to, and read from, the registers themselves.

%% file: sections/05-eval-peripheral.tex
\begin{table*}
    \centering
    \caption{Results Peripheral Device Verification}\label{tab:crossym-journal:eval:normal}
    \resizebox{\linewidth}{!}{
        \begin{tabular}{rcc rrrrrrc r rrrrrrc rr}
            \toprule
            &&& \multicolumn{6}{c}{\textbf{\crossym{}}} && \multicolumn{8}{c}{\textbf{\hwklee{}}} \\
            \cline{4-10} \cline{12-20}
            \multicolumn{3}{c}{\multirow{2}{*}{\textbf{Test}}} & \multicolumn{2}{c}{\textbf{Paths}} & \multirow{2}{*}{\textbf{Time}} & \multirow{2}{*}{\textbf{Solver}} & \multirow{2}{*}{\textbf{Memory}} & \multirow{2}{*}{\textbf{Queries}} & \multirow{2}{*}{\textbf{Comment}} && \multicolumn{2}{c}{\textbf{Paths}} & \multirow{2}{*}{\textbf{Time}} & \multirow{2}{*}{\textbf{Solver}} & \multirow{2}{*}{\textbf{Memory}} & \multirow{2}{*}{\textbf{Queries}} & \multirow{2}{*}{\textbf{Comment}} & \multicolumn{2}{c}{\textbf{\% Increase}}\\
            \multicolumn{3}{c}{} & compl. & partial &&&&&& & compl. & partial &&&&& & Time & Mem \\

            \hline

            \textbf{1} & \parbox[t]{2mm}{\multirow{5}{*}{\rotatebox[origin=c]{90}{PLIC}}} & T1 &
            168 & 0 & 581.33 & 73.32\% & 335.20 & 15893 & \checkmark &&
            165 & 0 & 998.29 & 69.29\% & 641.95 & 15779 & \checkmark          & \color{red}71.73 & \color{red}91.51 \\

            \textbf{2} & & T2 &
            0 & 530 & 798.59 & 80.18\% & 319.53 & 24235 & E1 &&
            0 & 530 & 689.44 & 90.66\% & 607.43 & 12606 & E1  & \color{blue!20!black!30!green}-13.67 & \color{red}90.10 \\

            \textbf{3} & & T3 &
            3677 & 6386 & TO & 90.56\% & 2010.03 & 903046 & (\checkmark) &&
            2217 & 5912 & TO & 90.77\% & 2091.61 & 655979 & (\checkmark)  & 0.00 & \color{red}4.06 \\

            \textbf{4} & & EQ2 &
            37 & 90 & TO & 96.99\% & 361.24 & 32504 & E1 &&
            465 & 415 & 77996.14 & 93.98\% & 1413.11 & 787434 & E1      & \color{blue!20!black!30!green}-9.73 & \color{red}291.18 \\

            \textbf{5} & & EQ3 &
            31 & 154 & TO & 97.11\% & 398.41 & 63828 & E2 &&
            371 & 374 & TO & 94.19\% & 1273.24 & 821189 & E2         & 0.00 & \color{red}219.58 \\

            \hline

            \textbf{6} & \parbox[t]{2mm}{\multirow{3}{*}{\rotatebox[origin=c]{90}{GCD}}} & \gls{rtl} &
            57 & 1564 & TO & 99.61\% & 260.23 & 4701 & STO,\sout{INF} &&
            53 & 1217 & TO & 97.89\% & 569.70 & 95577 & STO,\sout{INF} & 0.00 & \color{red}118.92 \\

            \textbf{7} & & \gls{tlm} &
            54 & 2204 & TO & 98.89\% & 298.58 & 25661 & E3,STO &&
            54 & 2197 & TO & 99.34\% & 474.73 & 4686 & E3,STO & 0.00 & \color{red}59.00 \\

            \textbf{8} & & EQ &
            428 & 6287 & TO & 95.57\% & 468.29 & 126779 & E3,\sout{INF} &&
            138 & 1516 & TO & 96.14\% & 649.35 & 149376 & E3,\sout{INF} & 0.00 & \color{red}38.66 \\

            \hline

            \textbf{9} & \parbox[t]{2mm}{\multirow{3}{*}{\rotatebox[origin=c]{90}{Hash}}} & \gls{rtl} &
            0 & 1 & 122.37 & 98.11\% & 245.70 & 2 & STO &&
            0 & 1 & 122.30 & 98.18\% & 436.38 & 2 & STO & \color{blue!20!black!30!green}-0.06 & \color{red}77.61 \\

            \textbf{10} & & \gls{tlm} &
            1 & 0 & 0.44 & 0.00\% & 234.66 & 0 & \checkmark &&
            1 & 0 & 1.34 & 0.00\% & 423.88 & 0 & \checkmark & \color{red}204.55 & \color{red}80.64 \\

            \textbf{11} & & EQ &
            0 & 1 & 122.69 & 97.85\% & 250.59 & 2 & STO &&
            0 & 1 & 122.72 & 97.85\% & 450.61 & 2 & STO & \color{red}0.02 & \color{red}79.82 \\

            \hline

            \textbf{12} & \parbox[t]{2mm}{\multirow{3}{*}{\rotatebox[origin=c]{90}{Map}}} & \gls{rtl} &
            256 & 0 & 148.13 & 7.19\% & 270.54 & 256 & \checkmark &&
            256 & 0 & 130.02 & 26.99\% & 479.47 & 766 & \checkmark & \color{blue!20!black!30!green}-12.23 & \color{red}77.23 \\

            \textbf{13} & & \gls{tlm} &
            1 & 0 & 0.51 & 0.00\% & 233.44 & 0 & \checkmark &&
            1 & 0 & 1.37 & 0.00\% & 424.62 & 0 & \checkmark & \color{red}168.63 & \color{red}81.90 \\

            \textbf{14} & & EQ &
            256 & 0 & 150.33 & 7.18\% & 275.58 & 256 & \checkmark &&
            256 & 0 & 153.50 & 23.95\% & 501.00 & 766 & \checkmark & \color{red}2.11 &\color{red} 81.80 \\
            \bottomrule
        \end{tabular}
    }
    \\[0.5em]
    \begin{minipage}{\linewidth}
        \centering
        Time in \unit{\second}, Average Memory in \unit{\mebi\byte}. TO=\qty{24}{\hour} Timeout, STO=Solver Timeout, E[1-3]=Error[1-3], \sout{INF}=Infinite Loop not found
    \end{minipage}
\end{table*}

%% file: sections/05-eval.tex
\section{Evaluation}\label{sec:crossym-journal:eval}
To evaluate our two tools \crossym{} and \hwklee{}, we concerned ourselves with six research questions:
\begin{enumerate}[leftmargin=1.2cm]
    \item[\textbf{RQ1}] Can our tools fully explore the functionality of different peripherals?
    \item[\textbf{RQ2}] Can our tools fully explore the bus interfaces of different peripherals?
    \item[\textbf{RQ3}] Can our tools offer effective bug finding?
    \item[\textbf{RQ4}] How does the array minimization technique influence the performance?
    \item[\textbf{RQ5}] How does the complexity of the integrated \systemc{} kernel influence the performance?
    \item[\textbf{RQ6}] How do our approaches scale to established benchmarks?
\end{enumerate}
We base our experiments on the verification of four different peripherals available at RTL and TLM: a \riscv{} \gls{plic}, and modules to compute the \gls{gcd}, a hash, and value mappings for a list, respectively.
We believe these modules represent a reasonable choice as they cover a range of different properties.
The hash and \gls{gcd} are their module's only output and are calculated with a fixed-length loop and an input-dependant one, respectively.
The value mapping is implemented by a \gls{mmio} peripheral, sharing memory between the bus and the peripheral itself instead of having a separate address space.
The \gls{plic} realises a comparatively more complex logic that follows the SiFive FE310 specification~\cite{sifive-hifive1-manual}.
Following our two proposed tools \crossym{} and \hwklee{}, we have two setups.
Either the \systemc{} kernel or the symbolic execution engine have to be modified, as discussed in \cref{sec:crossym-journal:main:challenges}.
Both symbolic execution engines were configured the same way.
As the \gls{smt} solver, STP~\cite{stp-solver} is used.
In order to achieve results within an acceptable time frame, we imposed additional limits:
an execution runs a maximum of \qty{24}{\hour}, and can use at most \qty{4000}{\mebi\byte} of memory.
A single solver query can take at most \qty{120}{\second}, so that the runtime is not dominated by few solver queries.
This is 4 times the time limit applied in~\cite{klee-multismt}, an in-depth study on \gls{smt} solvers for symbolic execution.
The search strategy for choosing execution states during the exploration is \gls{bfs}.
For \hwklee{}, the Array Minimization technique discussed in \cref{sec:crossym-journal:main:array-min} is activated.
It prevents large arrays in \gls{smt} solver queries, which would slow the execution down.
All tests were performed on a Linux Ubuntu 22.04 with an Intel Xeon Gold 6240 with 2.6 GHz.

\subsection{Peripheral Device Verification}\label{sec:crossym-journal:eval:normal}
To investigate our tools' appropriateness towards a full peripheral verification, we applied them to all four peripherals.
These are available in an \gls{rtl} and \gls{tlm} implementation, resulting in three verification scenarios for each peripheral: standalone \gls{rtl}, standalone \gls{tlm}, and cross-level.
The \gls{plic} is based on the SiFive FE310 SoC~\cite{sifive-hifive1-manual}.
It manages global interrupts, for example from Input/Output devices, and notifies the targets, usually Hardware Threads.
The interrupt order is determined by their individual priorities, as well as the per-target priority threshold.
In our setup, it supports a maximum of 52 interrupt sources, and a maximum priority of 7.
The \gls{tlm} test cases for the \gls{plic} are not discussed in this section, but in \cref{sec:crossym-journal:eval:perf-array,sec:crossym-journal:eval:perf-time}.
This is because they already featured in the case study of the \symsysc{} paper~\cite{pascal-symsys}, and we can therefore use them to compare our tools' performance to the state-of-the-art.
In this section, the \gls{plic} is only discussed at \gls{rtl}.
The standalone tests for the \gls{rtl} \gls{plic} consider the options which influence the firing of an interrupt.
Test \#1 triggers one input with a symbolic interrupt number.
Test \#2 adds to this a symbolic priority and priority threshold.
Test \#3 checks the correct interrupt order given two symbolic interrupts, each with a symbolic priority.
Tests \#4 \& \#5 mirror tests  \#2 \& \#3, but applied to the cross-level verification.
For the remaining four peripherals, the approach results in three test cases each: \gls{rtl} standalone, \gls{tlm} standalone, and cross-level.
The \gls{gcd} module calculates the greatest common divisor of two input values using the subtraction-based Euclidean algorithm.
For the tests, these two inputs are made symbolic, and in the standalone test, the result is compared to the division-based Euclidean algorithm.
The Hash module calculates a hash value of two inputs, which are again made symbolic.
In the standalone test, the result is compared against the hash function itself.
The Map module applies a function to map eight input values to eight output values.
All eight inputs are made symbolic, in addition to the mapping value.
In the standalone test, the result is compared against the mapping function itself.
The symbolic inputs are mostly written directly into the peripheral registers, as discussed in \cref{sec:crossym-journal:main:symsig}.
Writing symbolic data over the communication interfaces is evaluated separately in \cref{sec:crossym-journal:eval:bus}.

\Cref{tab:crossym-journal:eval:normal} shows the results for each test case.
It is split into three parts, first defining the test case, and then giving the results of our two tools.
For each tool, six columns show the number of completely and partially explored paths, the duration (in \unit{\second}), the percentage of time spent in the \gls{smt} solver, the average memory usage (\unit{\mebi\byte}), and the amount of queries passed to the solver.
A seventh column summarises the results.
The last two columns show \hwklee{}' time and memory increase in comparison with \crossym{} in percent, calculated using the ration of the two values.
In the following, different aspects of these results are discussed, starting with the \gls{duv} errors, followed by timeouts observed in both tools, and finally the differences between the tools.

\input{sections/05-eval-bus}

\subsubsection{Peripheral Implementation Errors}
The \gls{plic} is one of two peripherals in which functional errors were found.
In the \gls{rtl} implementation, the relationship between the priority and the threshold was inverse (\textbf{E1}):
the interrupt was fired if the priority was less than or equal to the threshold, instead of if it was greater.
This error was found through both the standalone test case and in the cross-level comparison.
Additionally, the cross-level comparison shows that the two \glspl{plic} handle priority values greater than the user-defined maximum differently (\textbf{E2}).
The other \gls{duv} with a functional error was the \gls{gcd} module.
Both the standalone \gls{tlm} and the cross-level test identified wrong results stemming from an internal unsigned-signed conversion (\textbf{E3}).
However, one known bug, which triggers an infinite loop, is not identified by either tool (\textbf{\sout{INF}}).
A corresponding path is identified, but not completed, which means the error could only be found by manually executing all generated concrete test cases.
Missing this error is an example for the main issues arising from the overall timeout and solver timeouts.
\subsubsection{Tool-independent Timeouts}
The (solver) timeouts (\textbf{(S)TO}) are the most reoccurring problem during the verification.
In the \gls{gcd} and hash modules, their reasons can be further examined by the queries passed to the solver.
For the \gls{gcd} module, the internal loop depends on the symbolic input variables, which means a significantly larger state space, and, as time goes on, more complex symbolic conditions.
In the Hash module, the internal loop runs a fixed number of times, limiting the state space.
However, the problem lies in the comparison of the resulting hash values.
The calculation constructs a complex symbolic expression, which is structured differently for the \gls{tlm} and \gls{rtl} implementations.
If their structures is equal, the comparison is fast, as can be seen in test case \#10.
The test case can finish in less than two seconds, because the chosen reference method is the same as the one called internally by the \gls{tlm} implementation.
For the \gls{plic}, both cross-level test cases and one standalone test generally do not finish under the \qty{24}{\hour} \textbf{TO}.
The priority cases (\Cref{tab:crossym-journal:eval:normal}, tests \#3 and \#5) suffer from a similar problem as the \gls{gcd} module.
The threshold test case (\#4) shows exemplary that the cross-level scenario takes longer, because two peripheral implementations have to be explored.
Nonetheless, the cross-level tests do find the aforementioned errors.
Finally, the Map module is the only \gls{duv} that reliably finishes without any timeouts.
This can be explained with the relatively simple implementation, featuring neither complex expressions nor symbolic loop conditions.
In the \gls{tlm} standalone test (\#13), the same phenomenon as for the Hash peripheral in test \#10 can be observed.

\subsubsection{Tool Performance Comparison}
Apart from these universal timeouts, the tools' performance also differs in a number of test cases.
\hwklee{} has to execute the slower \systemc{} startup, which becomes apparent in overall short test cases such as \#10 or \#13.
Generally, both tools are somewhat similar in speed, though \hwklee{} does have to use more memory to keep up with \crossym{}.
This originates mostly from the array minimization technique.
In overall slow tests, such as the cross-level ones, even small performance differences can have an impact.
Test \#4 is a case where \hwklee{} benefits from a few percentage points, and finishes in under \qty{24}{\hour}, while \crossym{} does not.

Through these standalone and cross-level verifications, we were able to evaluate four different peripherals, and further investigate how our two tools compare with each other.
Both tools found implementation errors and extensively explored the \glspl{duv} within a reasonable time frame.
Our evaluation also allows us to formulate concrete questions regarding the verification of some peripherals.
Further research has to identify solutions for complex symbolic expressions (see tests \#9 and \#11), large state spaces (see tests \#6 to \#8) and time-consuming complex paths (see tests \#4 and \#5).
One especially important aspect of this are unbounded loops, which are quintessential to hardware models.
While our current approach relies entirely on restrictions defined by the engineer in the test cases, more structured approaches are possible.
Examples are loop summaries~\cite{symex-ueberblick}, or the state pruning and memory smudging techniques that already proved efficient for firmware~\cite{firmware-loops}.
During this section, we largely ignored the communication interfaces in favor of a faster runtime, as discussed in \cref{sec:crossym-journal:main:symsig}.
Since the correct communication of data is paramount to correct results, these interfaces are evaluated in the following.

\subsection{Peripheral Communication Interface Verification}\label{sec:crossym-journal:eval:bus}
One integral part of peripherals is the additional logic required to connect them to the bus.
Even a flawless peripheral logic cannot alleviate issues resulting from flawed data transactions to/from the peripheral.
Therefore, this section focuses on the communication interfaces of the four peripherals at both \gls{tlm} and \gls{rtl}.
At each level, the interfaces are structurally similar for all four peripherals.
Their integration differs with regards to amount and addresses of registers, and behavior on transaction.
We implemented a total of 16 tests, reading and writing data over each interface.
The address, data length, and in the case of writing, the data itself, are made symbolic.
After the transaction, the peripheral's register values are checked.
The \gls{tlm} \gls{plic}'s tests are based on the \symsysc{} case study~\cite{pascal-symsys}.
However, the original tests only checked if any errors are triggered during the transaction.
They did not consider the actual effects of the transaction.
We extended this with assertions targeting the register values.

\input{sections/05-eval-mutation}
\Cref{tab:crossym-journal:eval:bus} shows the results for each test case.
It is split into three parts, first defining the test case, and then giving the results of our two tools.
For each tool, six columns show the number of completely and partially explored paths, the duration (in \unit{\second}), the percentage of time spent in the \gls{smt} solver, the average memory usage (\unit{\mebi\byte}), and the amount of queries passed to the solver.
A seventh column summarises the results.
The last two columns show \hwklee{}' time and memory increase in comparison with \crossym{} in percent, calculated using the ration of the two values.
The \gls{tlm} tests identified various errors.
Most of these (\textbf{E1,E3,E4}) are known from~\cite{pascal-symsys}, and are assertion errors for invalid addresses or data length.
Additionally, a new ``out of bound'' error is found in the \gls{tlm} Map (\textbf{E2}), triggered by writing to an invalid address.
All \gls{rtl} interfaces correctly complete transactions without triggering any errors or falsely modifying registers.
These results were generated quickly, with most tests completing in less than ten minutes.
One outlier is the \gls{tlm} \gls{plic}'s write test, which takes more than the \qty{24}{\hour} time limit.
This behavior was already documented for the state-of-the-art \symsysc~\cite{pascal-symsys}, and is related to the complex logic that is triggered during certain write transactions.
Finally, an important observation is the higher runtime for \hwklee{}.
While the percentage increase seems drastic, this mostly originates from the short absolute runtime.
Constant one-time efforts, such as a longer startup of the original \systemc{}, represent a larger part of this runtime.
\hwklee{} nonetheless takes at most around \qty{15}{\minute}, except for the outlier discussed previously.
However, in the timeout case, the slower runtime means \hwklee{} misses \textbf{E4}, and does not even identify the corresponding path.
The memory usage also increases, though less drastically than the runtime.
A decrease in memory usage is visible for the timeout case, where fewer paths are discovered, and with them, fewer opportunities for memory-intense queries.
Both of these trends mostly mirror the previous observations.

In summary, we demonstrate the efficient verification of communication interfaces by both of our tools.
This supports our choice of separating the verification of functional logic and communication interfaces, as outlined in \cref{sec:crossym-journal:main:symsig}.
When verified separately, both can offer valuable insights.
To better understand our approaches' bug-finding capabilities, the following section deals with controlled mutations systematically inserted into the \glspl{duv}.

\subsection{Effective Mutation-Based Bug Finding}\label{sec:crossym-journal:eval:mutations}
While our tools were sometimes not able to fully explore the \glspl{duv}, errors were found early on.
Identifying errors fast is an important verification goal to minimise the danger from the potentially unexplored parts of the peripheral.
In order to evaluate how suited our available verification scenarios and tools are for finding errors, we inserted mutations into the four \glspl{duv}.
We replaced all operators with ones from the same class, e.g. arithmetic.
In a single line of code, we considered changes of both a single and multiple operators.
Of the resulting mutants, we chose over 300 as a representative subset that changed the behavior e.g. by accessing the wrong/no register, or calculation errors.
While we aimed for 50 mutants for each peripheral and abstraction level, this depends on the code's complexity.
Both the \gls{gcd} and Map modules at \gls{tlm} did not offer enough opportunities for operator modifications.
For each mutant, the functional test cases discussed in \cref{sec:crossym-journal:eval:normal} were executed.
Since we only want to find out if the error can be found, instead of fully exploring the \gls{duv}, we can abort after the first error.
For this, we had to fix the previously discussed bugs.
Additionally, we lowered the maximum runtime to \qty{15}{\minute}.

\Cref{tab:crossym-journal:eval:mutations} shows our tools' results of the bug finding.
It is split into three parts, first defining the tests, and then giving the results of our two tools.
The tests are categorised by the mutated \gls{duv}, and the amount of mutants as well as of executed standalone (T) or cross-level (EQ) test cases.
For each tool three columns show the amount of alive and killed mutants, and the percentage of killed mutants.
``Killed'' in this context refers to the mutants that were successfully identified as not matching the desired behavior.
For \hwklee{}, a fourth column compares the result with \crossym{}, defined as the difference in percentage points.
The last column summarises the overall result.
The mutants were commonly killed due to failed assertions, overshift, memory, ``unreachable instruction'' or ``divide by zero'' errors.
However, some test cases ran into the timeout without detecting an error.
These timeouts can have numerous origins.
First, the observations from \Cref{tab:crossym-journal:eval:normal} have an influence.
If both tools found an identical, or nearly identical, number of paths in \Cref{tab:crossym-journal:eval:normal}, the bug finding performance is generally similar (\textbf{O5}).
If one tool found an overall drastically lower number of paths, the bug finding performance may also be worse (\textbf{O2}), especially given the shorter time.
Additionally, cross-level tests are generally slower.
This often translates to fewer killed mutants, in comparison to the standalone scenario (\textbf{O3}).
However, these are not golden rules, because other factors can also play a role.
For example, the second possible timeout cause are the paths through the test cases, and the resulting symbolic execution state tree.
The structure of this tree can be decisive in whether a mutant is found, especially given the \gls{bfs} search strategy.
This structure is influenced by the code of the \gls{duv}, the \systemc{} kernel, and the test case itself.
One example of this can be seen for \crossym{} and the \gls{gcd} \gls{rtl} mutants, where the cross-level scenario actually performs better (\textbf{O4}).
The \gls{gcd} \gls{tlm} code, which is executed first, creates paths that favor the detection of the mutants early.
If the order of the peripherals is switched, and the mutated \gls{rtl} peripheral is executed first, the statistics become comparable to those of the standalone scenario.
Note that the \gls{gcd} \gls{rtl} does not have the same effect for the \gls{tlm} mutants.
At the same time, \hwklee{} has a high success rate for the \gls{gcd} \gls{rtl} mutants, regardless of scenario.
Here, the original \systemc{} implementation can offer paths that lead to errors fast (\textbf{O4}).
This can also be observed for the Hash \gls{tlm} mutants.
It is notable that this means that the absolute number of paths discovered in the general verification (\Cref{tab:crossym-journal:eval:normal}) does not necessarily translate to fast bug finding success.
For \gls{gcd}, \hwklee{} had fewer paths, and for Hash, both tools found the same.
\input{sections/05-eval-perf-array}
\input{sections/05-eval-perf-symsys}
Therefore, in addition to the number of paths, the relevance of a path for a given bug counts, especially under time constraints.
A third timeout cause can be the number of separate test cases in a verification scenario.
Specifically, the \gls{plic} is the only peripheral where each verification scenario consists of multiple tests.
A mutant's behavior may well be triggered by only a subset of the executed tests.
Similarly, a final timeout cause is relevant for the Hash module.
Here, the general verification was mostly burdened by a complex query for comparing the calculation results.
This can also be observed in the bug finding, where a number of mutants related to the calculation itself exit with a solver timeout, without being killed (\textbf{O6}).
These solver timeouts occur independently of verification scenario or tool.
In contrast to the first three peripherals, all Map module mutants were always found.
This matches the observation that the verification of this module usually only takes less than \qty{15}{\min}.
Consequently, bugs can also reliably be identified in such a timeframe.

With the mutation-based evaluation, our approach proved capable of finding over 300 mutants fast ($<$\qty{15}{\minute}).
The results unearth how the statistics of the general verification relate to bug finding.
Complex interactions can be observed between verification runtimes and scenarios, as well as state trees created by the structure of peripheral, \systemc{} kernel, and testbench.
Generally, \hwklee{} can offer a slightly better performance for the bug finding.
It is aided in this effort by the array minimization, which was applied in all experiments discussed so far.
In the following section, we demonstrate the drastic influence it has on \hwklee{}' overall performance.

\subsection{Performance Comparison: Array Minimization}\label{sec:crossym-journal:eval:perf-array}
As part of \hwklee{}, we proposed the Array Minimization technique, which reduces arrays to the entries relevant for a given expression.
This aims for an overall lower solver time, as array queries are generally slow for large object sizes.
To examine the performance effects of this gamble, we utilize the \gls{tlm} \gls{plic} case study from the \symsysc{} paper~\cite{pascal-symsys}.
This already serves as basis for the performance comparison to \symsysc{} itself, as discussed in \Cref{sec:crossym-journal:eval:perf-time}.
For the \gls{tlm} \gls{plic}, the standalone test cases mirror those of the \gls{rtl} discussed above.
The last two cases handle \gls{tlm} transactions, focusing on reading and writing, respectively.
Note that these two are executed in their original form, without the additional register value checks discussed in \Cref{sec:crossym-journal:eval:bus}.
For the following two sections, the \gls{plic} is configured with the original setting from the \symsysc{} paper, namely a maximum of 64 interrupt sources, and a maximum priority of 32.
We executed all five tests for \hwklee{} both without and with Array Minimization activated.

\Cref{tab:crossym-journal:eval:optimisation} shows the run metrics for both versions.
It is split into three parts, first defining the test case, then giving the performance of \hwklee{} without, and with Array Minimization.
For each, four columns list the complete and partial executed paths, the time (in \unit{\second}), and average memory usage (in \unit{\mebi\byte}).
The last columns show the time/memory improvements with Array Minimization in percent, calculated using the ration of the two values.
Both found the same errors, which are not listed for brevity.
They relate to assertions that have no place in production environments.
The drastic improvement introduced by the array minimization is immediately obvious.
With it, \hwklee{} reduces the timeouts to two test cases.
In one of these, test \#3, it nonetheless finds drastically more paths, and completely explores most of them.
For tests \#1 and \#2, timeouts can not just be avoided, but results are generated in only around \qty{1}{\hour} and \qty{9}{\hour}, respectively.
These benefits come at the cost of a higher memory usage.
Tests \#4 and \#5 stand out, as they are the only ones where array minimization leads to a worse performance.
In \#4, the version with array minimization ends up taking slightly longer.
Similarly, in \#5, array minimization leads to drastically fewer completely or partially discovered paths.
Consequently, fewer opportunities for memory-intense queries arise.
Both tests focus on a \gls{tlm} transaction, without additional assertions in the test itself.
Fewer queries involve the large \gls{plic} object.
Therefore, the overhead of array minimization does not result in noticeable benefits.
These observations help understand the drastic improvement offered by the array minimization.
This section, alongside the previous ones, allowed us to evaluate the benefits of our contributions for a cross-level verification of \systemc{} peripherals.
They additionally allowed for a deeper insight into the differences between \crossym{} and \hwklee{}.
Finally, the following section will compare our tools to the current state-of-the-art, \symsysc~\cite{pascal-symsys}.

\subsection{Performance Comparison: State-of-the-Art}\label{sec:crossym-journal:eval:perf-time}
Both our replacement kernel and our extended symbolic execution tool offer more features than the state-of-the-art.
This can influence the performance, which is important because the verification should find results with a reasonable resource usage.
To compare our tools to the state-of-the-art, we repeated the case study of the \symsysc{} paper~\cite{pascal-symsys}.
In addition to \crossym{} and \hwklee{}, we consider the combination of the modified \systemc{} implementation from \crossym{} and the modified symbolic execution engine from \hwklee{}.
While \hwklee{}' threading approximation is not necessary for modified \systemc{}, the array minimization could influence the performance.
The test cases and settings are the same as for \Cref{sec:crossym-journal:eval:perf-array}.
\Cref{tab:crossym-journal:eval:performance} shows the run metrics for both versions.
It is split into four parts, first defining the test case, then giving the performance of \symsysc~\cite{pascal-symsys}, \crossym{}, \hwklee{}, and the combination of both modified components.
For each, four columns list the complete and partial executed paths, the time (in \unit{\second}), and average memory usage (in \unit{\mebi\byte}).
For our three versions, two columns list the time/memory increase in comparison to \symsysc~\cite{pascal-symsys}.
This is calculated using the ration of the two values.
All four versions found the same errors, which are not listed for brevity.
This supports that, for the original \gls{tlm} features, our tools offer the same bug finding capabilities.
However, in addition, they also offer more features, allowing a wider range of verification scenarios.
Surprisingly, \hwklee{} is mostly faster than \symsysc{}, even though it supports the more complex original \systemc{}.
Key for this is the array minimization, the effects of which are also discussed above.
It is not applied in \symsysc{}, and even \symsysc{}'s smaller replacement kernel creates objects large enough to highlight this issue.
However, \hwklee{} shows a higher memory usage, which is caused only in part by the array minimization.
Even without it, the more complex objects created by the original \systemc{} also lead to a higher memory usage than the replacement kernels.
\crossym{} comes with only a slight trade-off in comparison to \symsysc{}.
The two replacement kernels are very similar for the \gls{tlm} features that are evaluated here, which means similar results can be expected.
Their main difference is the support for \systemc{} \gls{rtl} features, which, if unused, only adds to object sizes and number of branch instructions.
One exception is test \#3, where \crossym{} reports lower memory, although both approaches exceed the time limit.
\crossym{} does not advance as far into the state space within the same time.
Therefore, it has fewer opportunities for memory-intense queries.
Test \#5 does not have the same decrease, likely due to the number of paths.
In the full explorations (tests \#1,\#2,\#4), these are identical for both replacement kernels.
For test \#3, \crossym{} found 65 paths total, with 31 fully explored, while \symsysc{} has 127 fully explored ones and an additional 64 incomplete.
Thus, \crossym{} is far behind regarding the state space.
Likewise, for test \#5, \crossym{} and \symsysc{} both found 264464 paths, of which \symsysc{} fully explored 15186, i.e. only 912 more.
Thus, the relative difference is not as staggering, which reflects in the memory increase.

Combining the two modified components underlines these observations.
It can be drastically faster than all three other tools by using both array minimization and the simpler \systemc{} implementation.
However, having both modifications instead of only one limits its practical applicability, as modifications can represent possible weak points in the verification process.
Additionally, the strict separation between \crossym{} and \hwklee{} throughout our paper allows us to show the equivalence between the modified and unmodified version of both components.

Overall, our proposed tools are not only comparable to the state-of-the-art, but additionally offer more advanced features to include \gls{rtl} modules and enable cross-level verification with symbolic execution.
\crossym{} remained especially close, with a temporal increase of under 10\%, and a memory increase of under 30\%.
In comparison to these two, \hwklee{} benefits from modifications to the symbolic execution engine.
These make its runtime competitive with the replacement kernels, while supporting the original \systemc{}.

\subsection{Scalability Investigation}
In the previous sections, our four peripherals allowed for a detailed investigation into challenges and advantages of, as well as differences between our tools.
Using peripherals from the \mbox{S2C} benchmark~\cite{s2c}, we can further investigate how our tools scale for established SystemC modules.
For each module, we apply symbolic inputs and compare the outputs to the expected results.
We removed the \qty{120}{\second} solver time limit due to the overall performance issues discussed below.
\input{sections/05-eval-s2c}
\Cref{tab:crossym-journal:eval:s2c} shows the results for both tools.
It is split into three parts, first defining the test case, then giving the performance of \crossym{} and \hwklee{}.
For each, four columns list the complete and partial executed paths, the time (in \unit{\second}), and average memory usage (in \unit{\mebi\byte}).
For \hwklee{}, the last column lists the memory increase in comparison to \crossym{}.
This is calculated using the ration of the two values.

The main difference between both tools' performance is the number of paths.
\hwklee{} completes more paths, but finds less overall.
The original \systemc{} implementation can create early restrictions on the symbolic variables, which results in paths that are easier to solve.
Nonetheless, \hwklee{} is slower in making overall progress, given the drastically lower number of paths.
However, both tools have timeouts for all modules, and hit the memory limit in nearly all cases.
This highlights how the limitations discussed in \Cref{sec:crossym-journal:eval:normal} transfer to established modules.
While our tools allow for a fundamental comparison between the different possible approaches for symbolic execution of \systemc{} peripherals, the open challenges are defined by our evaluation.
Different approaches for these challenges are possible, for example summarising paths or loops, or tailoring search strategies to \systemc{} modules.
The current performance makes it unfeasible to center other evaluation aspects, such as the bug finding, around the S2C peripherals.
Test cases that can be fully explored offer unique, more detailed insight into the differences between our tools.

\subsection{Summary}
This evaluation examined our two tools under three different criteria, 1) peripheral verification, 2) bug finding, and 3) performance.
For most of the evaluation, we considered four peripherals, each available at both \gls{rtl} and \gls{tlm}.
For the verification, separating peripheral functionality and communication interfaces proved effective.
Regarding the communication interfaces, an efficient verification was possible, with the exploration generally terminating in less than \qty{15}{\minute}.
Regarding the functionality, both tools identified implementation errors and extensively explored the \glspl{duv} within a reasonable time frame.
The second criteria was fast bug finding, which we examined with a mutation-based evaluation.
Over 300 mutants could be identified fast ($\leq$\qty{15}{\minute}).
The largest successfully explored \gls{duv}, regardless of tool or scenario, was the Map peripheral, where \crossym{} visited at most 1015 branch instructions, and \hwklee{} at most 1831.
We observed complex interactions between different influences on the runtime, and therefore on the bug finding success.
These highlight an advantage of our versatile approaches, because it can be difficult to predict which verification scenario is most beneficial for a \gls{duv}.
Furthermore, the last criteria, performance, unearthed that the Array Minimization is highly beneficial for \hwklee{}, enabling a reasonable runtime.
In comparison to the state-of-the-art, both our tools offer a comparable performance, despite their vastly more diverse feature set.
In comparison amongst each other, our evaluation showed that \hwklee{} generally requires more memory and time than \crossym{}.
For shorter test cases, the longer initial setup can become apparent.
However, \hwklee{} supports the original, unmodified \systemc{} kernel and \gls{duv}.
An important additional takeaway is the functional equivalence between these unmodified versions and the modified ones.
Finally, we observed some limitations for our tools.
\crossym{} supports only a constrained subset of \systemc{} features, since it relies on a replacement implementation of the kernel.
This limitation is not as prominent in \hwklee{}, which integrates the original \systemc{} implementation.
However, as outlined in \Cref{sec:crossym-journal:main:app-op}, both approaches focus on verifying a single peripheral, and do not support asynchronous waiting.
Further limitations were highlighted by our evaluation.
As is common in symbolic execution, both tools encounter the state space explosion~\cite{symex-ueberblick}, e.g. in the \gls{plic} and \gls{gcd} modules.
This can lead to runtimes above \qty{24}{\hour}, or solver runtimes above \qty{120}{\second}.
Additionally, complex symbolic expressions can also lead to solver runtimes above \qty{120}{\second}, as observed in the Hash module.
In an additional evaluation on peripherals from the S2C benchmark, these limitations mean that our approaches do not yet scale well to larger modules.
Nonetheless, we demonstrated that our tools have practical use, and typically find errors fast.

%% file: sections/05-eval-bus.tex
\begin{table*}
    \centering
    \caption{Results Peripheral Bus Interface Verification}\label{tab:crossym-journal:eval:bus}
    \resizebox{\linewidth}{!}{
        \begin{tabular}{rccc rrrrrrc r rrrrrrc rr}
            \toprule
            &&&& \multicolumn{6}{c}{\textbf{\crossym{}}} && \multicolumn{8}{c}{\textbf{\hwklee{}}} \\
            \cline{5-11} \cline{13-21}
            \multicolumn{4}{c}{\multirow{2}{*}{\textbf{Test}}} & \multicolumn{2}{c}{\textbf{Paths}} & \multirow{2}{*}{\textbf{Time}} & \multirow{2}{*}{\textbf{Solver}} & \multirow{2}{*}{\textbf{Memory}} & \multirow{2}{*}{\textbf{Queries}} & \multirow{2}{*}{\textbf{Comment}} && \multicolumn{2}{c}{\textbf{Paths}} & \multirow{2}{*}{\textbf{Time}} & \multirow{2}{*}{\textbf{Solver}} & \multirow{2}{*}{\textbf{Memory}} & \multirow{2}{*}{\textbf{Queries}} & \multirow{2}{*}{\textbf{Comment}} & \multicolumn{2}{c}{\textbf{\% Increase}}\\
            \multicolumn{4}{c}{} & compl. & partial &&&&&& & compl. & partial &&&&& & Time & Mem \\

            \hline

            \textbf{1} & \parbox[t]{2mm}{\multirow{4}{*}{\rotatebox[origin=c]{90}{\gls{plic}}}} & \multirow{2}{*}{RTL} & Read &
            113 & 0 & 39.90 & 37.58\% & 287.70 & 492 & \checkmark &&
            113 & 0 & 288.94 & 13.53\% & 637.57 & 494 & \checkmark & \color{red}624.16 & \color{red}121.61 \\

            \textbf{2} & & & Write &
            255 & 0 & 230.44 & 78.41\% & 312.14 & 6454 & \checkmark &&
            255 & 0 & 780.54 & 50.71\% & 661.11 & 6461 & \checkmark & \color{red}238.72 & \color{red}111.80 \\

            \cline{3-21}

            \textbf{3} & &\multirow{2}{*}{TLM} & Read &
            970 & 198 & 74.41 & 241.07\% & 241.07 & 3300 & E1 &&
            970 & 198 & 342.28 & 55.35\% & 462.12 & 3300 & E1 & \color{red}359.99 & \color{red}91.70 \\

            \textbf{4} & & & Write &
            11903 & 165832 & TO & 96.45\% & 3357.08 & 493023 & E1, E3, E4 &&
            686 & 616 & TO & 99.29\% & 1073.70 & 912137 & E1,E3 & 0.00 & \color{blue!20!black!30!green}-68.02 \\

            \hline

            \textbf{5} & \parbox[t]{2mm}{\multirow{4}{*}{\rotatebox[origin=c]{90}{\gls{gcd}}}} & \multirow{2}{*}{RTL} & Read &
            6 & 0 & 1.81 & 14.56\% & 241.49 & 9 & \checkmark &&
            6 & 0 & 5.69 & 8.34\% & 455.78 & 11 & \checkmark          & \color{red}214.36 & \color{red}88.74 \\

            \textbf{6} & & & Write &
            15 & 0 & 4.45 & 32.49\% & 248.24 & 47 & \checkmark &&
            15 & 0 & 288.04 & 91.69\% & 472.96 & 3313 & \checkmark & \color{red}6372.81 & \color{red}90.53 \\

            \cline{3-21}

            \textbf{7} & & \multirow{2}{*}{TLM} & Read &
            25 & 6 & 3.44 & 71.46\% & 240.18 & 62 & E1 &&
            25 & 6 & 11.01 & 29.27\% & 446.16 & 62 & E1 & \color{red}220.06 & \color{red}85.76 \\

            \textbf{8} & & & Write &
            15 & 8 & 6.94 & 87.34\% & 242.87 & 165 & E1,E3 &&
            15 & 8 & 405.41 & 95.81\% & 454.61 & 10849 & E1,E3 & \color{red}5741.64 & \color{red}87.18 \\

            \hline

            \textbf{9} & \parbox[t]{2mm}{\multirow{4}{*}{\rotatebox[origin=c]{90}{Hash}}} & \multirow{2}{*}{RTL} & Read &
            6 & 0 & 1.81 & 14.90\% & 241.55 & 9 & \checkmark &&
            6 & 0 & 5.76 & 8.32\% & 455.78 & 11 & \checkmark          & \color{red}218.23 & \color{red}88.69 \\

            \textbf{10} & & & Write &
            15 & 0 & 4.28 & 30.23\% & 248.28 & 42 & \checkmark &&
            15 & 0 & 183.96 & 89.36\% & 472.33 & 2048 & \checkmark & \color{red}4198.13 & \color{red}90.24 \\

            \cline{3-21}

            \textbf{11} & & \multirow{2}{*}{TLM} & Read &
            25 & 6 & 3.45 & 71.58\% & 240.24 & 62 & E1 &&
            25 & 6 & 11.15 & 28.81\% & 446.12 & 62 & E1 & \color{red}223.19 & \color{red}85.70 \\

            \textbf{12} & & & Write &
            15 & 8 & 7.21 & 88.59\% & 242.93 & 165 & E1, E3 &&
            15 & 8 & 391.10 & 95.68\% & 454.53 & 10849 & E1,E3 & \color{red}5324.41 & \color{red}87.10 \\

            \hline
            \textbf{13} & \parbox[t]{2mm}{\multirow{4}{*}{\rotatebox[origin=c]{90}{Map}}} & \multirow{2}{*}{RTL} & Read &
            9 & 0 & 2.47 & 17.37\% & 242.72 & 15 & \checkmark &&
            9 & 0 & 7.88 & 9.73\% & 460.32 & 17 & \checkmark          & \color{red}219.03 & \color{red}8.38 \\

            \textbf{14} & & & Write &
            18 & 0 & 5.49 & 37.72\% & 247.82 & 70 & \checkmark &&
            18 & 0 & 1011.46 & 96.43\% & 480.35 & 19426 & \checkmark & \color{red}18323.68 & \color{red}93.83 \\

            \cline{3-21}

            \textbf{15} &  & \multirow{2}{*}{TLM} & Read &
            35 & 8 & 5.93 & 57.24\% & 241.88 & 86 & E1 &&
            35 & 8 & 14.58 & 29.83\% & 448.53 & 86 & E1 & \color{red}145.87 & \color{red}85.43 \\

            \textbf{16} & & & Write &
            28 & 10 & 14.08 & 81.85\% & 245.79 & 302 & E1,E2 &&
            28 & 10 & 737.87 & 96.47\% & 459.48 & 17766 & E1,E2 & \color{red}5140.55 & \color{red}86.94 \\
            \bottomrule
        \end{tabular}
    }
    \\[1em]
    \begin{minipage}{\linewidth}
        \centering
        Time in \unit{\second}, Average Memory in \unit{\mebi\byte}. TO=\qty{24}{\hour} Timeout, E[1-4]=Error[1-4]
    \end{minipage}
\end{table*}

%% file: sections/05-eval-mutation.tex
\begin{table*}
    \centering
    \caption{Mutation Test Results (O[0-6] = Observation[0-6])}\label{tab:crossym-journal:eval:mutations}
    \begin{tabular}{ccccr rrr r rrrr c}
        \toprule
        \multicolumn{2}{c}{\multirow{2}{*}{\textbf{\gls{duv}}}} & \multirow{2}{*}{\textbf{Mutants}} & \multicolumn{2}{c}{\multirow{2}{*}{\textbf{Tests}}} & \multicolumn{3}{c}{\textbf{\crossym{}}} && \multicolumn{4}{c}{\textbf{\hwklee{}}} & \multirow{2}{*}{\textbf{Comment}} \\
        \cline{6-8}\cline{10-13}
        &&&&& \textbf{Alive} & \textbf{Killed} & \textbf{\%} && \textbf{Alive} & \textbf{Killed} & \textbf{\%} & \textbf{\% Increase} & \\

        \hline

        \parbox[t]{2mm}{\multirow{4}{*}{\rotatebox[origin=c]{90}{\gls{plic}}}} & \multirow{2}{*}{\gls{rtl}} & \multirow{2}{*}{50} & T & 150 &
        46 & 104 & 69.33 &&
        4 & 146 & 97.33 & \color{blue!20!black!30!green}28.00 & O1 \\
        &&& EQ & 100 &
        100 & 0 & 0.00 &&
        25 & 75 & 75.00 & \color{blue!20!black!30!green}75.00 & O1,O3 \\
        \cline{2-14}
        & \multirow{2}{*}{\gls{tlm}} & \multirow{2}{*}{50} & T & 150 &
        40 & 110 & 73.33 &&
        37 & 113 & 75.33 & \color{blue!20!black!30!green}2.00 & O5,O1 \\
        &&& EQ & 100 &
        54 & 46 & 46.00 &&
        27 & 73 & 73.00 & \color{blue!20!black!30!green}27.00 & O1,O3 \\

        \hline

        \parbox[t]{2mm}{\multirow{4}{*}{\rotatebox[origin=c]{90}{GCD}}} & \multirow{2}{*}{\gls{rtl}} & \multirow{2}{*}{50} & T & 50 &
        14 & 36 & 72.00 &&
        2 & 48 & 96.00 & \color{blue!20!black!30!green}24.00 & O4 \\
        &&& EQ & 50 &
        1 & 49 & 98.00 &&
        2 & 48 & 96.00 & \color{red}-2.00 & O4 \\
        \cline{2-14}
        & \multirow{2}{*}{\gls{tlm}} & \multirow{2}{*}{22} & T & 22 &
        8 & 14 & 63.64 &&
        8 & 14 & 63.64 & 0.00 & O5 \\
        &&& EQ & 22 &
        11 & 11 & 50.00 &&
        15 & 7 & 31.81 & \color{red}-18.19 & O2,O3\\

        \hline

        \parbox[t]{2mm}{\multirow{4}{*}{\rotatebox[origin=c]{90}{Hash}}}  & \multirow{2}{*}{\gls{rtl}} & \multirow{2}{*}{50} & T & 50 &
        28 & 22 & 44.00 &&
        28 & 22 & 44.00 & 0.00 & O5,O6 \\
        &&& EQ & 50 &
        28 & 22 & 44.00 &&
        28 & 22 & 44.00 & 0.00 & O5,O6\\
        \cline{2-14}
        & \multirow{2}{*}{\gls{tlm}} & \multirow{2}{*}{50} & T & 50 &
        34 & 16 & 32.00 &&
        34 & 16 & 32.00 & 0.00 & O5,O6 \\
        &&& EQ & 50 &
        41 & 9 & 18.00 &&
        39 & 11 & 22.00 & \color{blue!20!black!30!green}4.00 & O4-O6 \\

        \hline

        \parbox[t]{2mm}{\multirow{4}{*}{\rotatebox[origin=c]{90}{Map}}}  & \multirow{2}{*}{\gls{rtl}} & \multirow{2}{*}{50} & T & 50 &
        0 & 50 & 100.00 &&
        0 & 50 & 100.00 & 0.00 & \checkmark \\
        &&& EQ & 50 &
        0 & 50 & 100.00 &&
        0 & 50 & 100.00 & 0.00 &  \checkmark \\
        \cline{2-14}
        & \multirow{2}{*}{\gls{tlm}} & \multirow{2}{*}{35} & T & 35 &
        0 & 35 & 100.00 &&
        0 & 35 & 100.00 & 0.00 &  \checkmark \\
        &&& EQ & 35 &
        0 & 35 & 100.00 &&
        0 & 35 & 100.00 & 0.00 & \checkmark \\
        \bottomrule
    \end{tabular}
\end{table*}

%% file: sections/05-eval-perf-array.tex
\begin{table}
    \setlength{\tabcolsep}{2.5pt}
    \centering
    \caption{Performance Comparison \hwklee{} with/without Array Minimization}\label{tab:crossym-journal:eval:optimisation}
    \resizebox{\linewidth}{!}{
    \begin{tabular}{c rrrr r rrrr rr}
        \toprule
        & \multicolumn{4}{c}{\textbf{\hwklee{} - Vanilla}} && \multicolumn{6}{c}{\textbf{\hwklee{} - Array Minimization}} \\
        \cline{2-5}\cline{7-12}
        \multirow{2}{*}{} & \multicolumn{2}{c}{\textbf{Paths}} & \multirow{2}{*}{\textbf{Time}} & \multirow{2}{*}{\textbf{Mem}} && \multicolumn{2}{c}{\textbf{Paths}} & \multirow{2}{*}{\textbf{Time}} & \multirow{2}{*}{\textbf{Mem}} & \multicolumn{2}{c}{\textbf{\% Increase}} \\
        & compl. & partial &&&& compl. & partial &&& Time & Mem \\

        \hline
        \textbf{1} & 0 & 5 & TO & 448.55 &&
        63 & 1 & 3744.13 & 489.77 & \color{blue!20!black!30!green}-95.67 & \color{red}9.19 \\

        \textbf{2} & 1 & 6 & TO & 453.76 &&
        189 & 0 & 32022.28 & 860.14 & \color{blue!20!black!30!green}-62.94 & \color{red}89.55 \\

        \textbf{3} & 1 & 14 & TO & 457.06 &&
        154 & 68 & TO & 1129.74 & 0.00 & \color{red}147.18 \\

        \hline
        \textbf{4} & 970 & 198 & 221.43 & 459.47 &&
        970 & 198 & 253.46 & 459.46 & \color{red}14.46 & 0.00 \\

        \textbf{5} & 2733 & 261731 & TO & 6279.18 &&
        1793 & 207273 & TO & 5039.35 & 0.00 & \color{blue!20!black!30!green}-19.75 \\
        \bottomrule
    \end{tabular}
    }
    \\[1em]
    \begin{minipage}{\linewidth}
        \centering
        Time in \unit{\second}, Average Memory in \unit{\mebi\byte}. TO=\qty{24}{\hour} Timeout
    \end{minipage}
\end{table}

%% file: sections/05-eval-perf-symsys.tex
\begin{table*}
    \setlength{\tabcolsep}{2.5pt}
    \centering
    \caption{Performance Comparison}\label{tab:crossym-journal:eval:performance}
    \resizebox{\linewidth}{!}{
        \begin{tabular}{c rrrr r rrrrrr r rrrrrr r rrrrrr}
            \toprule
            & \multicolumn{4}{c}{\textbf{\symsysc~\cite{pascal-symsys}}} &&
            \multicolumn{6}{c}{\textbf{\crossym{}}} &&
            \multicolumn{6}{c}{\textbf{\hwklee{}}} &&
            \multicolumn{6}{c}{\textbf{Modified \systemc{} \& Symbolic Execution}} \\
            \cline{2-5}\cline{7-12}\cline{14-19}\cline{21-26}
            \multirow{2}{*}{} & \multicolumn{2}{c}{\textbf{Paths}} & \multirow{2}{*}{\textbf{Time}} & \multirow{2}{*}{\textbf{Mem}} &&
            \multicolumn{2}{c}{\textbf{Paths}} & \multirow{2}{*}{\textbf{Time}} & \multirow{2}{*}{\textbf{Mem}} & \multicolumn{2}{c}{\textbf{\% Increase}} &&
            \multicolumn{2}{c}{\textbf{Paths}} & \multirow{2}{*}{\textbf{Time}} & \multirow{2}{*}{\textbf{Mem}} & \multicolumn{2}{c}{\textbf{\% Increase}} &&
            \multicolumn{2}{c}{\textbf{Paths}} & \multirow{2}{*}{\textbf{Time}} & \multirow{2}{*}{\textbf{Mem}} & \multicolumn{2}{c}{\textbf{\% Increase}} \\
            & compl. & partial && &&
            compl. & partial &&& Time & Mem &&
            compl. & partial &&& Time & Mem &&
            compl. & partial &&& Time & Mem \\

            \hline
            \textbf{1} & 63 & 1 & 7988.27 & 280.35 &&
            63 & 1 & 8592.80 & 335.26 & \color{red}7.57 & \color{red}19.59 &&
            63 & 1 & 3744.13 & 489.77 & \color{blue!20!black!30!green}-53.13 & \color{red}74.70 &&
            63 & 1 & 436.28 & 249.10 & \color{blue!20!black!30!green}-94.54 & \color{blue!20!black!30!green}-11.15 \\

            \textbf{2} & 189 & 0 & 71788.69 & 461.99 &&
            189 & 0 & 74733.22 & 599.53 & \color{red}4.10 & \color{red}29.77 &&
            189 & 0 & 32022.28 & 860.14 & \color{blue!20!black!30!green}-55.39 & \color{red}86.18 &&
            189 & 0 & 1467.35 & 272.59 & \color{blue!20!black!30!green}-97.96 & \color{blue!20!black!30!green}-41.00 \\

            \textbf{3} & 127 & 64 & TO & 406.01 &&
            31 & 34 & TO & 291.29 & 0.00 & \color{blue!20!black!30!green}-28.26 &&
            154 & 68 & TO & 1129.74 & 0.00 & \color{red}178.25 &&
            584 & 124 & TO & 883.45 & 0.00 & \color{red}117.59 \\

            \hline
            \textbf{4} & 970 & 198 & 103.58 & 191.89 &&
            970 & 198 & 112.37 & 239.95 & \color{red}8.49 & \color{red}25.04 &&
            970 & 198 & 253.46 & 459.46 & \color{red}144.70 & \color{red}139.44 &&
            970 & 198 & 186.52 & 245.38 & \color{red}80.07 & \color{red}27.87 \\

            \textbf{5} & 15186 & 249278 & TO & 3374.77 &&
            14274 & 250190 & TO & 3715.45 & 0.00 & \color{red}10.09 &&
            1793 & 207273 & TO & 5039.35 & 0.00 & \color{red}49.32 &&
            1475 & 24525 & TO & 1286.16 & 0.00 & \color{blue!20!black!30!green}-61.89 \\
            \bottomrule
        \end{tabular}
    }
    \\[1em]
    \begin{minipage}{\linewidth}
        \centering
        Time in \unit{\second}, Average Memory in \unit{\mebi\byte}. TO=\qty{24}{\hour} Timeout
    \end{minipage}
\end{table*}

%% file: sections/05-eval-s2c.tex
\begin{table}
    \setlength{\tabcolsep}{2.5pt}
    \centering
    \caption{Results Scalability Evaluation}\label{tab:crossym-journal:eval:s2c}
    \resizebox{\linewidth}{!}{
        \begin{tabular}{rc rrrr r rrrrr}
            \toprule
            && \multicolumn{4}{c}{\textbf{\crossym{}}} && \multicolumn{5}{c}{\textbf{\hwklee{}}} \\
            \cline{3-6} \cline{8-12}
            \multicolumn{2}{c}{\multirow{2}{*}{\textbf{Test}}} & \multicolumn{2}{c}{\textbf{Paths}} & \multirow{2}{*}{\textbf{Time}} & \multirow{2}{*}{\textbf{Mem}} && \multicolumn{2}{c}{\textbf{Paths}} & \multirow{2}{*}{\textbf{Time}} & \multirow{2}{*}{\textbf{Mem}} & \multirow{2}{*}{\textbf{\% Incr.}}\\
            & & compl. & partial & & && compl. & partial &&& \\

            \hline
            \textbf{1} & ADPCM &
            51 & 178493 & TO & \cellcolor{red!25!blue!25}3024.27 &&
            136 & 63990 & TO & \cellcolor{red!25!blue!25}3701.28 & \color{red}80.07 \\

            \textbf{2} & AES &
            0 & 186348 & TO & \cellcolor{red!25!blue!25}2726.47 &&
            0 & 40496 & TO & \cellcolor{red!25!blue!25}3400.49 & \color{red}24.72 \\

            \textbf{3} & Ave8 &
            597 & 153625 & TO & \cellcolor{red!25!blue!25}2678.83 &&
            917 & 52516 & TO & \cellcolor{red!25!blue!25}3410.40 & \color{red}27.31 \\

            \textbf{4} & FIR &
            0 & 24765 & TO & 763.11 &&
            0 & 4669 & TO & 683.25 & \color{blue!20!black!30!green}-10.47 \\

            \textbf{5} & QSort &
            15847 & 85015 & TO & 1674.14 &&
            7346 & 64206 & TO & \cellcolor{red!25!blue!25} 2928.13 & \color{red}74.90 \\

            \bottomrule
        \end{tabular}
    }
    \\[1em]
    \begin{minipage}{\linewidth}
        \centering
        Time in \unit{\second}, Average Memory in \unit{\mebi\byte}. TO=\qty{24}{\hour} Timeout,\\Purple Highlight = Memory Limit reached
    \end{minipage}
\end{table}

%% file: sections/06-conclusion.tex
\section{Conclusions}\label{sec:crossym-journal:ende}
We propose two opposite approaches for the verification of \systemc{} peripherals using modern symbolic execution tools.
Each addresses the incompatibility between unmodified \systemc{} kernel and unmodified state-of-the-art symbolic execution differently.
\crossym{} builds on the current state-of-the-art concept of using a replacement \systemc{} implementation, and is the first cross-level verification method of this kind.
This replacement kernel supports key features of both \gls{tlm} and \gls{rtl} modelling, while being optimised for symbolic execution of peripherals in isolation.
\hwklee{} (Symbolic Execution For Original SystemC) accommodates the original \systemc{} implementation by modifying the state-of-the-art symbolic execution engine KLEE~\cite{klee}.
It leverages the \systemc{} threading behaviour~\cite{systemc} to avoid the overhead associated with verifying general multi-threaded software.
As part of \hwklee{}, we propose the Array Minimization technique, which reduces arrays to the entries relevant for a given expression.
Independent of the tool, we address peripherals specifically by considering problems with writing symbolic values over the communication interfaces.
We evaluated \crossym{} and \hwklee{} under three different criteria, (1) peripheral verification, (2) bug finding, and (3) performance.
Both offered a versatile set of features effective in cross-level settings, while remaining competitive with the state-of-the-art in terms of performance.
Our evaluation highlighted the important considerations for a comparison between \crossym{} and \hwklee{}.
While \hwklee{} supports the unmodified \systemc{} kernel and \gls{duv}, it does generally require more memory and often more time.
For future work, we plan to increase the complexity that symbolic execution of SystemC modules can handle, using techniques targeting unbounded loops, similar paths, and search strategies.
Such work, as well as considerations of the asynchronous waiting, is detrimental for scenarios involving multiple peripherals, or more complex ones.
An additional important aspect is detecting concurrency bugs.